\documentclass[11pt]{article}
\usepackage{graphicx}
\usepackage{amssymb}
\usepackage[section]{placeins}

\newtheorem{theorem}{Theorem}

\def\proof{\noindent  {\underline {Proof}}. }
\def\square{ {\hfill \vrule height6pt width6pt depth1pt} \bigskip \medskip }
\def\cadremath#1{\vbox{\hrule\hbox{\vrule\kern8pt\vbox{\kern8pt
			\hbox{ {$\displaystyle #1 $ } }\kern8pt} 
			\kern8pt\vrule}\hrule}}

\def\C{\mathbb{C}}

\def\today{\number\day\space\ifcase\month\or Janvier \or F\'evrier \or  Mars
   \or Avril \or Mai \or Juin \or Juillet \or Ao\^ut \or Septembre \or Octobre
   \or Novembre \or D\'ecembre \fi\number \year}

\begin{document}

\centerline{Classical Bethe Ansatz and Normal Forms in the Jaynes-Cummings Model.}
\bigskip
\centerline{O. Babelon, B. Dou\c{c}ot}
\bigskip
\centerline{Laboratoire de Physique Th\'eorique et Hautes Energies\footnote{Tour 13-14, 4\` eme \'etage, Boite 126,
4 Place Jussieu,  75252 Paris Cedex 05.
% \url{http://www.lpthe.jussieu.fr}
 }, (LPTHE)}
\centerline{Unit\'e Mixte de Recherche UMR 7589}
\centerline{Universit\'e Pierre et Marie Curie-Paris 6 and CNRS; }
\bigskip

{\bf Abstract:} The Jaynes-Cummings-Gaudin model describes a collection of $n$ spins
coupled to an harmonic oscillator. It is known to be integrable, so one can define a moment
map which associates to each point in phase-space the list of values of the $n+1$ conserved
Hamiltonians. We identify all the critical points of this map and we compute the
corresponding quadratic normal forms, using the Lax matrix representation of the model.
The normal coordinates are constructed by a procedure which appears as a classical version
of the Bethe Ansatz used to solve the quantum model. We show that only elliptic or focus-focus
singularities are present in this model, which provides an interesting example
of a symplectic toric action with singularities. To explore these, we study in detail the degeneracies of the
spectral curves for the $n=1$ and $n=2$ cases. This gives a complete picture for the image
of the momentum map  (IMM) and the associated bifurcation diagram. For $n=2$ we found in particular 
some lines of rank 1 which lie, for one part, on the boundary of the IMM, where they behave like
an edge separating two faces, and which go, for another part, inside the IMM. 
  
\section{Introduction.}
Suppose we have a Hamiltonian system with Hamiltonian $H$ on a phase space of dimension $2n$ with coordinates $x_1, \cdots , x_{2n}$. 
A critical point $x^{(0)}_j$ is an equilibrium point 
$$
\left. {\partial H \over \partial x_j}\right\vert_{x_j=x^{(0)}_j} = 0 , \quad j=1,\cdots , 2n
$$
We can expand the Hamiltonian around such a point, $x_j=x_j^{(0)} + \delta x_j$, getting a quadratic form
$$
H = H^{(0)} + Q + \cdots, \quad Q=\sum_{i,j} Q_{ij} \delta x_i \delta x_j 
$$
The question then arises  to simplify or ``diagonalize'' the quadratic form $Q$. 
This problem is simple if we allow $Sl_{2n}(R)$ transformations of the $\delta x_j$, 
the matrix $Q_{ij}$ can then be diagonalized in the orthogonal group. 
However in the present case the natural group is the group of symplectic transformations and 
the diagonalization problem is not straightforward. Its solution has been given by Williamson~\cite{williamson36,Arnold97} 
in terms of six families of elementary quadratic forms 
on which we can decompose $H$. Each family corresponds to a certain pattern of eigenvalues of the associated linear Hamiltonian flow. There are two families for the 
eigenvalue 0, a third one for a pair of opposite real eigenvalues (hyperbolic flow), two more for a pair of purely imaginary eigenvalues (elliptic flow), and the last
one for quartets of complex eigenvalues of the form $\pm \alpha \pm i\beta$. Besides diagonal blocks, each family admits non-diagonal Jordan blocks of unlimited size.
%We shall not write here all the possible normal forms allowed by  Williamson's theorem, but in the sequel, the following ones will play an important role: 
%\begin{eqnarray*}
%H^{\mathrm{elliptic}} &=& \frac{\beta}{2} (p^2+q^2), \quad \mathrm{eigenvalues} \pm i\beta \\
%H^{\mathrm{hyperbolic}} &=& \alpha p q , \quad n=1, \quad \mathrm{eigenvalues} \pm \alpha \\
%H^{\mathrm{focus-focus}} &=& \alpha (p_1 q_1 + p_{2}q_{2}) + \beta (p_1 q_{2} - p_{2} q_1), \quad \mathrm{eigenvalues} \pm \alpha \pm i\beta 
%\end{eqnarray*}

The above problem of reducing $H$  to its  normal form is not completely trivial. 
Even less trivial is the case of an integrable system, in which case we have $n$ Poisson commuting Hamiltonians $H_i$. 
These functions define the moment map, from $R^{2n}$ to $R^{n}$, which sends $x$ to $(H_1(x),...,H_{n}(x))$. 
A critical point is a simultaneous critical point for all the $H_i$, in other words, it is a point at which the differential of the moment
map vanishes. Expanding the $H_i$ around such a point, we get $n$ Poisson commuting quadratic forms $Q_i$. 
For each of them, Williamson theorem applies, but since they commute, we
can reduce them to normal forms simultaneously. The simplest case is of a nondegenerate singularity, for which the 
Hamiltonian flows associated to the $n$ quadratic forms $Q_i$ realize a Cartan subalgebra of the Lie algebra $sp_{n}(R)$.
The fact that each $Q_i$ Poisson commutes with the others yields strong constraints on the form of the elementary blocks
which can appear in the  Williamson reduction of $Q_i$. It turns out that integrability excludes non-diagonal Jordan blocks, therefore
there exist canonical coordinates $q_1,\cdots q_n$, $p_1,\cdots, p_n$ such that the above quadratic forms can be reduced into the  
following quadratic polynomials:~\cite{Eliasson90}
\begin{eqnarray*}
P^{\mathrm{elliptic}}_j & = & p_j^2+q_j^2,\quad j=1,2, \cdots ,m_1 \\
P^{\mathrm{hyperbolic}}_j & = & p_j q_j , \quad j=m_1+1, \cdots m_1+m_2 \\
P^{\mathrm{focus-focus}}_j & = & p_j q_j + p_{j+1}q_{j+1}, \quad j=  m_1+m_2+1\cdots \\
P^{\mathrm{focus-focus}}_{j+1} & = & p_j q_{j+1} - p_{j+1} q_j ,\quad \cdots m_1+m_2 + 2 m_3
\end{eqnarray*}
where $m_1+m_2+2 m_3=n$. The triple $(m_1,m_2,m_3)$ is called the type of the critical point.
It encodes all the qualitative information on the system, in particular on the dimensions of 
fibers of the moment map in the vicinity of the corresponding critical value.
The above set of quadratic forms is refered to as the normal form for a non-degenerate critical point of an integrable system.

In the focus-focus case, $P^{\mathrm{focus-focus}}_j$ and $P^{\mathrm{focus-focus}}_{j+1} $ can be combined into a single complex quantity. Setting 
$$
B_j =p_j+i p_{j+1},\quad B_{j+1} = q_{j} +i q_{j+1}
$$
we have
$$
B_j \overline{B}_{j+1} = P^{\mathrm{focus-focus}}_j  - i P^{\mathrm{focus-focus}}_{j+1}
$$

In this paper we observe that the solution to this problem of the simultaneous reduction of the conserved Hamiltonians 
in the vicinity of critical points is the classical analog of Bethe Ansatz which is used to diagonalize simultaneously 
the {\em quantum} commuting Hamiltonians $H_i$. We demonstrate it on the particular example of the Jaynes-Cummings-Gaudin model~\cite{Dicke,JC,Gau83,YKA}, 
but the construction is clearly a general one. As a result we will show that, in this model, singularities are only of two types: elliptic and focus-focus.

The above critical points are the points where the moment map
$$
 R^{2n} \to R^{n} : \quad x \to (H_1(x),...,H_{n}(x))
$$
is of rank zero. Other manifolds where the rank of the moment map is not maximal are also of great interest. Their images in $R^{n}$ by the moment map constitute  the bifurcation diagram. As explained in particular by
Mich\`ele Audin~\cite{Audin96}, a powerful tool to 
construct this diagram for an integrable model 
is the correspondence between singularities of the moment map and
degeneracies of the associated spectral curve. We will show explicitely the results of this method
for the Jaynes-Cummings-Gaudin model with two and three degrees of freedom.
  
Another advantage of this model is that it admits one free parameter for each degree of freedom.
The types of the critical points therefore depend on these parameters.
In particular, it is possible to choose them in such a way that all the critical
points are of the elliptic type. In this case, the Hamiltonian flows of the conserved $H_i$  
can be used to define a torus action on phase space. 
A famous theorem states that, if we replace the $H_i$'s by
the corresponding action variables, 
the image of the moment map is a convex polytope~\cite{Atiyah82, GuilStern82}.
However, for most values of the parameters,
focus-focus singularities appear which provides examples of the more general ``almost toric'' action 
studied in~\cite{San05}, and the important phenomenon of ``monodromy'' discovered in~\cite{Duistermaat80}, 
preventing the existence of global action-angle variables, comes into play.

While the one-spin Jaynes-Cummings model was already re-discovered by mathematicians \cite{San10}, it seems that the results concerning the bifurcation diagrams of the two-spin Jaynes-Cummings model are new. The geometric richness of these diagrams undoubtedly  calls for further studies.

The paper is organized as follows. In section 2 we present the Jaynes-Cummings-Gaudin model and recall some basic facts about its Lax representation. In section 3 we study the critical points of the model and their normal forms  are computed in section 4 using a classical analog of Bethe Ansatz. In section 5 we discuss the monodromy. Section 6 is a short introduction to  separated variables and explains that the different strata of the bifurcation diagram correspond to the degeneracies of the spectral curve.  In sections 7 and 8 we study the bifurcation diagram for the one-spin and two-spin models respectively.

\section{The classical Jaynes-Cummings-Gaudin model.}

This model, where a collection of $n$ spins is coupled to a single harmonic oscillator,
has been used for more than fifty years in atomic physics to describe the interaction
of an ensemble of atoms with a mode of the quantized electromagnetic field~\cite{Dicke,JC,Gau83,YKA,BaDoCa09}.
It derives from the following Hamiltonian:
\begin{equation}
H= \sum_{j=1}^{n} (2 \epsilon_j +\omega)  s_j^z  + \omega \bar{b} b + 
\sum_{j=1}^{n} \left( \bar{b} s_j^-+ b s_j^+ \right)
\label{bfconspinclas}
\end{equation}
The $\vec{s}_j$ are spin variables, and $b,\bar{b}$ is a  harmonic oscillator.
The Poisson brackets read
\begin{equation}
\{ s_j^a , s_j^b \} = - \epsilon_{abc} s_j^c, \quad \{ b , \bar{b} \} = i
\label{poisson}
\end{equation}
The $\vec{s}_j$ brackets are degenerate. We fix the value of the Casimir functions
$$
\vec{s}_j \cdot \vec{s}_j = s^2
$$
Phase space has dimension $2(n+1)$. In the Hamiltonian we have used
$$
s_j^\pm = s_j^1 \pm i s_j^2
$$
which have Poisson brackets
$$
\{ s_j^z , s_j^\pm \} = \pm i s_j^\pm,\quad \{s_j^+,s_j^-\} = 2 i s_j^z
$$
The equations of motion read
\begin{eqnarray}
\dot{b} &=& -i {\partial H \over \partial \bar{b}}  =  -i \omega b - i  \sum_{j=1}^{n} s_j^-  \label{motionb} \\
\dot{s_j^z} &=& -i {\partial H \over \partial s_j^+} s_j^+ + i {\partial H \over \partial s_j^-} s_j^-  =  i  ( \bar{b} s_j^- - b s_j^+ ) \label{motionsz}\\
\dot{s_j^+} &=& i {\partial H \over \partial s_j^z} s_j^+ - 2 i {\partial H \over \partial s_j^-} s_j^z = i(2 \epsilon_j+\omega) s_j^+ -2i \bar{b} s_j^z \label{motions+} \\
\dot{s}_j^- &=&-i {\partial H \over \partial s_j^z} s_j^- + 2 i {\partial H \over \partial s_j^+} s_j^z =  -i(2 \epsilon_j +\omega) s_j^- +2i b s_j^z \label{motions-}
\end{eqnarray}

This is an integrable system. To see it we introduce the Lax matrices
\begin{eqnarray}
L(\lambda) &=& 2 \lambda \sigma^z + 2 (b \sigma^+ + \bar{b} \sigma^-) + \sum_{j=1}^{n} {\vec{s}_j \cdot \vec{\sigma} \over \lambda-\epsilon_j} 
\label{defLaxL} \\
M(\lambda) &=& -i\lambda \sigma^z  -i{ \omega\over 2}\sigma^z 
-i (b\sigma^+ + \bar{b} \sigma^-)
\label{defLaxM}
\end{eqnarray}
where $\sigma^a$ are the Pauli matrices.
$$
\sigma^x = \pmatrix{ 0 & 1 \cr 1 & 0}, \quad \sigma^y = \pmatrix{ 0 & -i \cr i & 0},\quad  \sigma^z = \pmatrix{1 & 0 \cr 0 & -1},\quad 
$$
and we have defined
$$
\sigma^\pm = {1\over 2} (\sigma^x \pm i \sigma^y), \quad [\sigma^z,\sigma^\pm ] = \pm  2 \sigma^\pm, \quad [\sigma^+, \sigma^-] = \sigma^z
$$
It is not difficult to check that the equations of motion are equivalent to the Lax equation
\begin{equation}
\dot{L}(\lambda) = [M(\lambda), L(\lambda) ]
\label{eqLax}
\end{equation}
Let
$$
L(\lambda) = \pmatrix{ A(\lambda) & B(\lambda) \cr C(\lambda) & -A(\lambda) }
$$
we have
\begin{eqnarray}
A(\lambda) &=& 2\lambda+ \sum_{j=1}^{n} {s_j^z \over \lambda - \epsilon_j } \label{defA}\\
B(\lambda) &=& 2b + \sum_{j=1}^{n} {s_j^- \over \lambda - \epsilon_j } \label{defB} \\
C(\lambda) &=& 2\bar{b} + \sum_{j=1}^{n} {s_j^+ \over \lambda - \epsilon_j }  \label{defC}
\end{eqnarray}
One has the simple  Poisson brackets
\begin{eqnarray}
\{A(\lambda), A(\mu) \} &=& 0 \label{AA}\\
\{B(\lambda), B(\mu) \} &=& 0 \label{BB}\\
\{C(\lambda), C(\mu) \} &=& 0 \label{CC}\\
\{A(\lambda), B(\mu) \} &=& {i\over \lambda - \mu} ( B(\lambda) - B(\mu) ) \label{AB}\\
\{A(\lambda), C(\mu) \} &=& -{i\over \lambda - \mu} ( C(\lambda) - C(\mu) ) \label{AC}\\
\{B(\lambda), C(\mu) \} &=& {2i\over \lambda - \mu} ( A(\lambda) - A(\mu) ) \label{BC}
\end{eqnarray}
One can rewrite these equations in the usual classical $r$-matrix form
$$
\{ L_1(\lambda) , L_2(\mu) \} =- i \left[ {P_{12}\over \lambda - \mu } , L_1(\lambda) + L_2(\mu) \right]
$$
where
$$
P_{12} = \pmatrix{ 1 & 0 & 0 & 0 \cr 0 & 0 & 1 & 0 \cr 0 & 1 & 0 & 0 \cr 0 & 0 & 0 & 1}
$$
It follows  immediately that $ \rm{Tr}\,(L^2(\lambda) ) = 2 A^2(\lambda) + 2 B(\lambda) C(\lambda)  $ Poisson commute for different values of the spectral parameter:
$$
\{ \rm{Tr}\,(L^2(\lambda_1) ), \rm{Tr}\,(L^2(\lambda_2) ) \} = 0
$$
Hence $\Lambda(\lambda)\equiv \frac{1}{2}\rm{Tr}\,(L^2(\lambda) ) $ generates Poisson commuting quantities.   One has
\begin{eqnarray}
\Lambda(\lambda)  =  {Q_{2n+2}(\lambda)\over \prod_j (\lambda-\epsilon_j)^2}=  4\lambda^2  + 4  H_{n+1}  + 
 2 \sum_{j=1}^{n}  
{H_j  \over \lambda - \epsilon_j}  +  \sum_{j=1}^{n} {s^2 \over ( \lambda - \epsilon_j)^2 }
\label{detL}
\end{eqnarray}
where the $(n+1)$ Hamiltonians $H_j$, $j=1,\cdots , n+1$  read
\begin{equation}
H_{n+1} =  b\bar{b} +  \sum_j s_j^z 
\label{Hn}
\end{equation}
and
\begin{equation}
H_j =  2\epsilon_j  s_j^z +  ( b s_j^+ + \bar{b} s_j^-)
+ \sum_{k\neq j} {s_j \cdot s_k \over \epsilon_j - \epsilon_k }, \quad j=1,\cdots , n
\label{Hj}
\end{equation}
One can easily verify that, indeed,  $\{ H_i, H_j \}=0$ for $i,j=1,\cdots, n+1$, hence the system is integrable.
The Hamiltonian eq.(\ref{bfconspinclas}) is
$$
H = \omega H_{n+1} + \sum_{j=1}^{n} H_j
$$

\section{Critical points}

The critical points are equilibrium points for all the Hamiltonians $H_j$, $j=1,\cdots, n+1$. At such points the derivatives with respect of all coordinates on phase space vanish. 
In particular, since
$$
{\partial H_{n+1}\over \partial \bar{b}} = b,\quad {\partial H_j\over \partial \bar{b}} =  s_j^-
$$
we see that the critical points must be located at
\begin{equation}
b=\bar{b} =0, \quad s_j^{\pm} = 0, \quad s_j^z = e_j s, \quad e_j = \pm 1
\label{static}
\end{equation}
When we expand around a configuration eq.(\ref{static}), all the quantities ($b$, $\bar{b}$, $s_j^+$, $s_j^-$) are first order, but $s_j^z$ is second order because
$$
s_j^z = e_j \sqrt{s^2-s_j^+ s_j^- } = s e_j -{e_j\over 2s} s_j^+ s_j^-  + \cdots, \quad e_j = \pm 1
$$
It is then simple to see that all first order terms in the expansions of the Hamiltonians $H_j$ vanish. Hence we have found  $2^n$ critical points.

\section{Normal Forms.}
\label{sec_Normal_Forms}

We want to expand the Hamiltonians $H_j$ around the equilibrium points eq.(\ref{static}) and write them  in normal form. Symbolically :
\begin{equation}
H_j = \sum_\alpha E_{j,\alpha}\; \bar{a}_\alpha a_\alpha
\label{Hjnormal}
\end{equation}
where $\bar{a}_\alpha, a_\alpha$ are independent harmonic oscillators (in the elliptic case). Note that 
if we quantize the above Hamiltonians in this approximation, their diagonalization  is immediate: 
the normal coordinates $a_\alpha$ become spectrum generating operators.
Their construction must therefore be  very much related  to the simultaneous diagonalization of the $H_j$. But the tool to solve this problem is well known: Bethe Ansatz. 

\bigskip

Inspired by this remark, we return to eqs.(\ref{defA} -- \ref{defC}) and eqs.(\ref{AA} -- \ref{BC}). Now, we have
$$
\left\{ {1\over 2}{\rm Tr}\; L^2(\lambda), C(\mu) \right\} = {2i\over \lambda-\mu} \Big( A(\lambda) C(\mu) - A(\mu) C(\lambda) \Big)
$$
When we expand around a critical point, the Hamiltonians are quadratic. Remark that $C(\mu)$ is first order and therefore the Poisson bracket in  left hand side is linear. Now $A(\lambda)$ is constant plus second order, so that in the right-hand side we can replace $A(\lambda)$ and $A(\mu)$ by their 
zeroth order expression :
$$
A(\lambda) \simeq a(\lambda) = 2\lambda + \sum_{j=1}^{n} {s e_j \over \lambda-\epsilon_j}
$$
and  we arrive at
\begin{equation}
\left\{ {1\over 2}{\rm Tr}\; L^2(\lambda), C(\mu) \right\} = {2i\over \lambda-\mu} \Big( a(\lambda) C(\mu) - a(\mu) C(\lambda) \Big)
\label{eq1}
\end{equation}
Going back to eq.(\ref{Hjnormal}), we see that the $a_\alpha$ are such that
\begin{equation}
\{ H_j, a_\alpha \} = E_{j,\alpha} a_\alpha
\label{eqnormal}
\end{equation}
Eq.(\ref{eq1})  will be precisely of the form of eq.(\ref{eqnormal}) if we can kill the unwanted term $C(\lambda)$. This is achieved by imposing the condition 
\begin{equation}
a(\mu)=0, \quad  \mbox{ ``Classical Bethe Equation''}
\label{classicalBethe}
\end{equation}
This is an equation of degree $n+1$ for $\mu$. Calling $\mu_i$ its solutions,  we construct in this way $n+1$ variables $C(\mu_i)$. Remark that by eq.(\ref{CC}),  they all commute
\begin{equation}
\{ C(\mu_i), C(\mu_j) \} = 0
\label{CiCj}
\end{equation}

 Since phase space has dimension $2(n+1)$ this is half what we need. To construct the conjugate variables, we consider eq.(\ref{BC}). In our linear approximation it reads
$$
\{ B(\mu_i), C(\mu_j)\} = {2i\over \mu_i-\mu_j} (a(\mu_i)-a(\mu_j) )
$$
If $\mu_i$ and $\mu_j$ are {\it different } solutions of eq.(\ref{classicalBethe}), then obviously
\begin{equation}
\{ B(\mu_i), C(\mu_j)\} =0, \quad \mu_i \neq \mu_j
\label{BiCj}
\end{equation}
If however $\mu_j=\mu_i$ then
\begin{equation}
\{ B(\mu_i), C(\mu_i)\} =  2i a'(\mu_i)
\label{BiCi}
\end{equation}
Finally, by eq.(\ref{BB}) we have
\begin{equation}
\{ B(\mu_i), B(\mu_j)\} = 0
\label{BiBj}
\end{equation}
Up to normalisation, we have indeed constructed canonical coordinates !

\bigskip

It is simple to express the quadratic Hamiltonians in theses coordinates:

\begin{equation}
{1\over 2}{\rm Tr}\; L^2(\lambda) = a^2(\lambda) + \sum_j { a(\lambda)\over a'(\mu_j) (\lambda-\mu_j) } B(\mu_j) C(\mu_j)
\label{benoit}
\end{equation}
This has the correct analytical properties in $\lambda$ and together with  the Poisson brackets eqs.(\ref{CiCj},\ref{BiCj}, \ref{BiCi}, \ref{BiBj})  we reproduce eq.(\ref{eq1}). Note that there is no pole at $\lambda=\mu_j$ because $a(\mu_j)=0$.  Expanding around $\lambda=\infty$ we get
$$
H_{n+1}= s\sum_k e_k + \sum_i {1\over 2 a'(\mu_i)} B(\mu_i) C(\mu_i)
$$
and computing the residue at $\lambda=\epsilon_j$, we find
$$
H_j= se_j\left[ 2\epsilon_j + \sum_k {se_k \over \epsilon_j-\epsilon_k}  \right] + \sum_i {1\over 2 a'(\mu_i)} { se_j \over \epsilon_j-\mu_i} B(\mu_i) C(\mu_i) 
$$

We can invert these formulae: devide eq.(\ref{benoit}) by $\lambda-\mu_j$ and take the residue at $\lambda=\mu_j$. Since $a(\mu_j)=0$ we get
$$
{1\over 2}{\rm Tr}\; L^2(\mu_j) =  B(\mu_j) C(\mu_j)
$$
or explicitly
$$
 B(\mu_j) C(\mu_j) = 4\mu_j^2 + 4 H_{n+1} + \sum_{k=1}^n {2H_k\over \mu_j-\epsilon_k} + \sum_{k=1}^n {s^2\over (\mu_j-\epsilon_k)^2}
 $$

\noindent
We can now make contact with the Williamson classification theorem.

\bigskip

If $\mu_j$ is {\em real} we have $B(\mu_j)=\overline{C(\mu_j)}$ and we set
$$
C(\mu_j) = \sqrt{ |a'(\mu_j) |} (p_j+i \epsilon_j  q_j), \quad B(\mu_j) = \sqrt{ |a'(\mu_j) |} (p_j-i\epsilon_j q_j )
$$
where $\epsilon_j = - {\rm sign}(a'(\mu_j))$ and $p_j, q_j$ are canonical coordinates. Then
eq.(\ref{BiCi}) is satisfied. Moreover 
$$
B(\mu_j) C(\mu_j) =  \vert a'(\mu_j) \vert (p_j^2+ q_j^2 ) 
$$
 i.e. we have an {\em elliptic} singularity.

\bigskip

If $\mu_j$ is {\em complex}, there is another root $\mu_{j+1}= \bar{\mu}_j$ which is its complex conjugate.
Then $\overline{B(\mu_j)}=C(\mu_{j+1})$. We introduce canonical coordinates $p_j,q_j,p_{j+1},q_{j+1}$ and  set
$$
B(\mu_j) = -ia'(\mu_j) (q_{j} + i q_{j+1}) ,\quad B(\mu_{j+1} )= p_j + i p_{j+1}
$$
$$
C(\mu_j) = p_j -i p_{j+1},\quad C(\mu_{j+1})= i a'(\mu_{j+1})  (q_j -i q_{j+1} )
$$
then eqs.(\ref{CiCj},\ref{BiCj},\ref{BiCi},\ref{BiBj}) are satisfied and
\begin{eqnarray*}
B(\mu_j)C(\mu_j) &=& -i a'(\mu_j) (P^{focus-focus}_j  +i P^{focus-focus}_{j+1} ) \\
B(\mu_{j+1})C(\mu_{j+1}) &=& i a'(\mu_{j+1}) (P^{focus-focus}_j  - i P^{focus-focus}_{j+1} ) 
\end{eqnarray*}
i.e. we have a {\em focus-focus} singularity.

\begin{figure}[hbtp]
\begin{center}
\frame{
\includegraphics[height= 5cm]{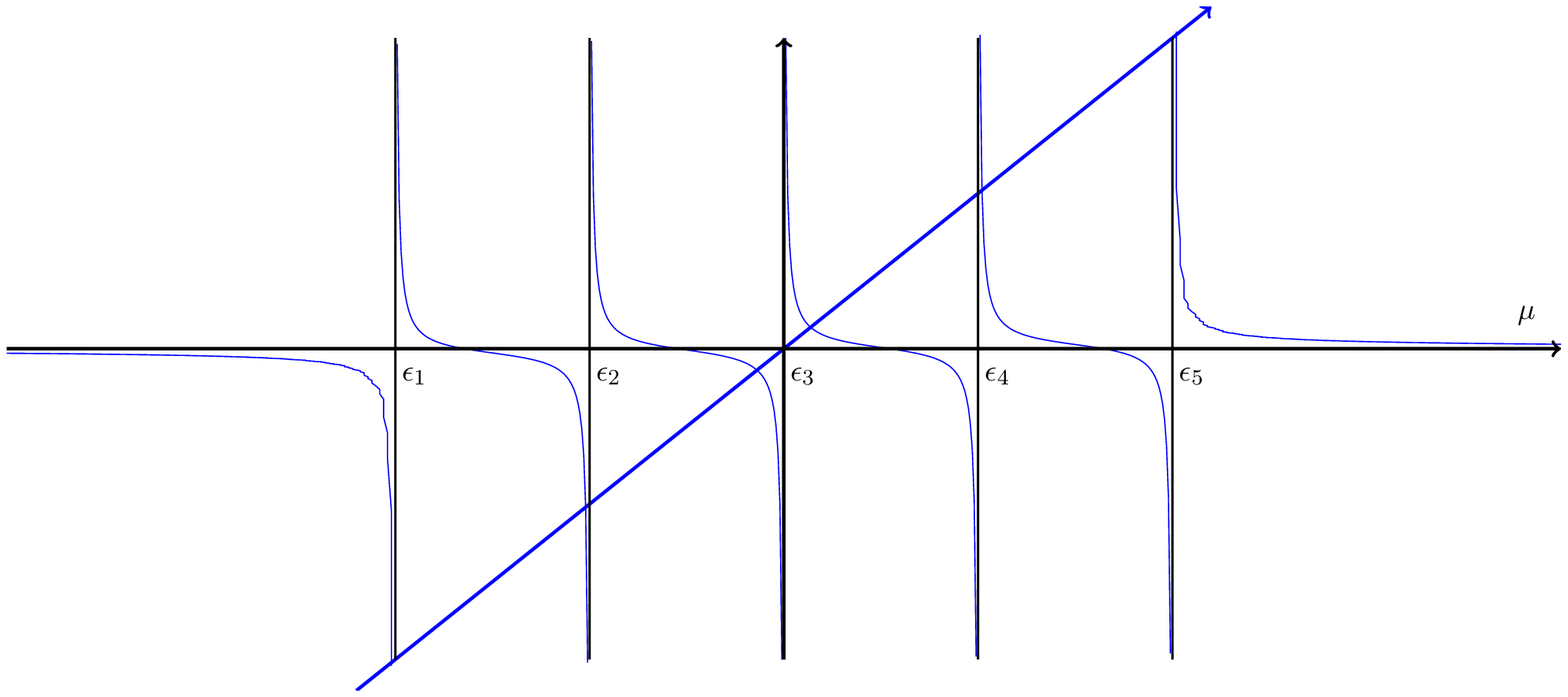} 
}
\caption{The solutions of eq.(\ref{eqE}) when  $e_j=-1$.}
\label{ej-1}
\end{center}
\nonumber
 \end{figure}

\bigskip
It remains to see when the classical Bethe roots are real and when they are complex. 
Let us assume all the spins are down: $e_j = -1$, $j=0,\cdots,n-1$.  The condition $a(\mu)=0$ reads
\begin{equation}
 \mu = {s\over 2}\sum_i {1\over \mu-\epsilon_i}
\label{eqE}
\end{equation}
The graph of the curves $y = \mu $ and $y = {s\over 2} \sum_{j=1}^{n} {1\over  \mu-\epsilon_j }$ are presented in Fig.[\ref{ej-1}]. In that case  we have $n+1$ {\em real} roots. The singularity is elliptic  meaning that this critical point is locally stable.
This is the situation that prevails when $\sum e_i <0$. 
\bigskip

\begin{figure}[hbtp]
\begin{center}
\frame{
\includegraphics[height= 5cm]{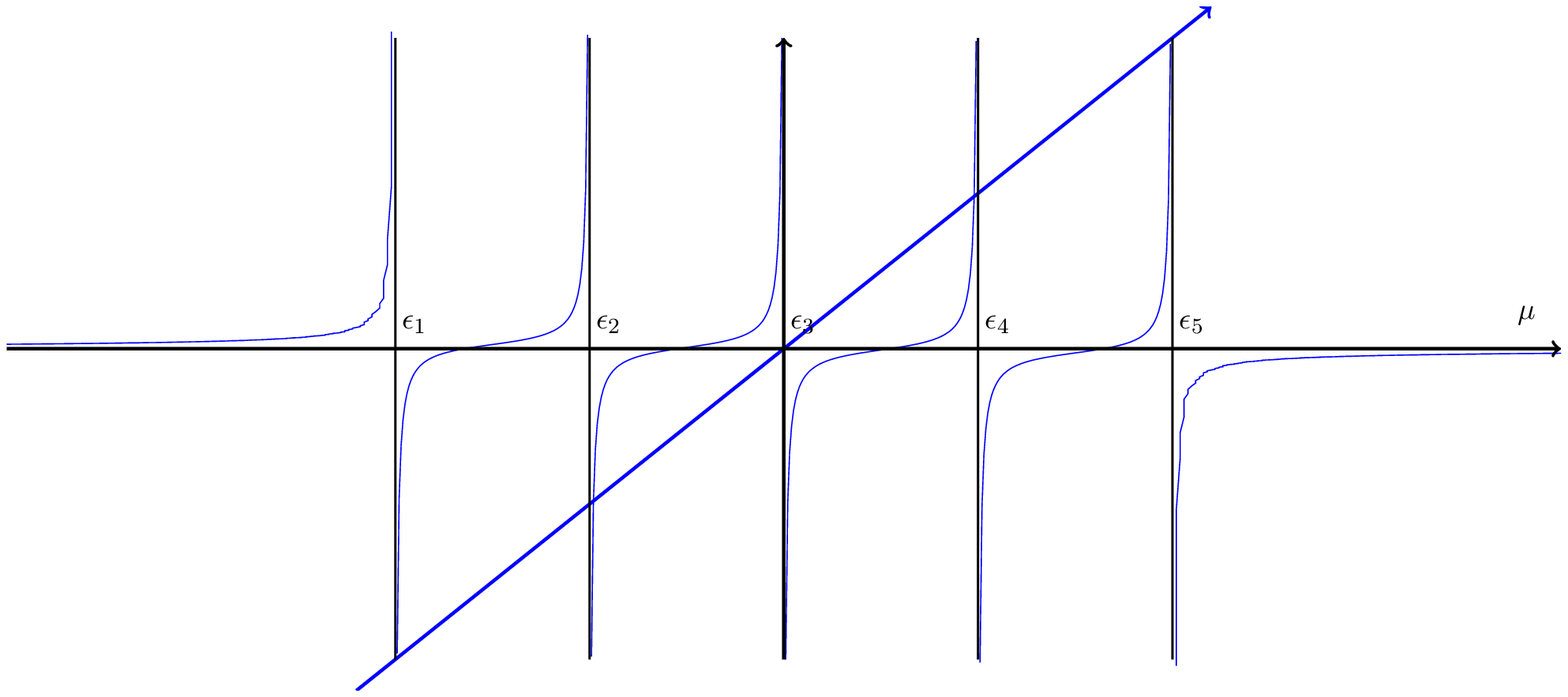} 
}
\caption{The solutions of eq.(\ref{eqE}) when $e_j= 1$.}
\label{ej+1}
\end{center}
\nonumber
 \end{figure}

Suppose now that $e_j = 1$, $j=0,\cdots,n-1$. The graph of the curves $y = \mu $ and $y = -{s\over 2} \sum_{j=1}^{n} {1\over  \mu-\epsilon_j }$ are presented in Fig~[\ref{ej+1}]. The situation is more complex, we can have  $n-1$ real roots and a {\em pair of complex conjugated roots},  or $n+1$ real roots, depending on the values of the $\epsilon_j$. 

This situation prevails when $\sum e_i > 0$.
When $\sum e_i = 0$, the driving parameter parameter is $\sum \epsilon_i e_i $.

\bigskip

Normal forms can be used to compute the dimension of the preimage of a singular point. Fixing the values of the conserved quantities
amount to fix  the values of the quadratic terms. For an elliptic term we have
$$
p_j^2 + q_j^2 = h_j
$$
and the preimage is a circle.  If $h_j=0$ this circle degenerates to a point. In the case of a focus-focus singularity we have
\begin{eqnarray*}
p_j q_j + p_{j+1} q_{j+1} &=& h_j \\
p_j q_{j+1} - p_{j+1} q_j &=& h_{j+1}
\end{eqnarray*}
Solving in $p_i$ for instance we find
$$
p_j = {h_j q_j + h_{j+1} q_{j+1} \over q_j^2+ q_{j+1}^2 }, \quad p_{j+1} = {h_j q_{j+1} - h_{j+1} q_j \over q_j^2+ q_{j+1}^2 }
$$
so we have a two dimensional preimage. If $h_j=h_{j+1}=0$, we have two planes $q_j=q_{j+1}=0$ or $p_j=p_{j+1}=0$ which intersect in one point. The preimage is a pinched torus.

\section{Monodromy.}
Let us go back to eq.(\ref{benoit}). We have
\begin{eqnarray}
{Q_{2n+2}(\lambda) \over \prod_j (\lambda-\epsilon_j)^2} &=&  a^2(\lambda) + \sum_j { a(\lambda)\over a'(\mu_j) (\lambda-\mu_j) } B(\mu_j) C(\mu_j)  \label{benoit2}
\end{eqnarray}
Of course
$$
a(\lambda) = 2{\prod_j(\lambda-\mu_j) \over \prod_j(\lambda-\epsilon_j)}
$$
We want to study the roots of the polynomial $Q_{2n+2}(\lambda)$ when we perform a small circle around the singularity in the space $H_i$ i.e. we want to examine the motion of the branch points of the spectral curve (to be defined in the next section, see eq.~(\ref{def_spectral_curve})) 
when we turn around a singularity. To zeroth order in the small deviations $b,\bar{b},s_{j}^{\pm}$ away from a critical point, the polynomials $B(\lambda)$
and $C(\lambda)$ vanish, whereas $A(\lambda)$ can be replaced by the fixed function $a(\lambda)$. Therefore:
$$
{Q_{2n+2}(\lambda) \over \prod_j (\lambda-\epsilon_j)^2} =  a^2(\lambda)
$$
and $Q_{2n+2}(\lambda)$ has double zeroes at the roots $\mu_j$ of the classical Bethe equation $a(\mu_j)=0$.
Let us now pick a phase-space point close to the critical point. The corresponding invariant polynomial $Q_{2n+2}(\lambda)$
is expressed in terms of the normal coordinates  $B(\mu_j)$, $C(\mu_j)$ through  eq.(\ref{benoit2})
Consider a  root $\mu_j$. The branch point is  moved at $\mu_j+\delta \mu_j$. Inserting into eq.(\ref{benoit2}) we get the equation
$$
(\delta \mu_j)^2 = -\left( {1\over a'(\mu_j)}\right)^2 B(\mu_j)C(\mu_j)
$$
so the leading variation of the root $\mu_{j}$ is due to the normal mode $B(\mu_j), C(\mu_j)$ only. The other modes contribute in subdominant terms. This is Krichever's result~\cite{Krich83}.

If the root $\mu_j$ is real, we see that that it splits  into a pair of complex conjugated roots, and the splitting is in the imaginary direction. To first order the deformation space is one dimensional, because, as we have seen, $B(\mu_j)=\overline{C(\mu_j)}$.

If the root $\mu_j$ is complex then $(\delta \mu_j)^2$ is complex. 
$$
(\delta \mu_j)^2 = -\left( {1\over a'(\mu_j)}\right)^2 B(\mu_j)C(\mu_j) = {i\over  a'(\mu_j)} (P^{focus-focus}_j  +i P^{focus-focus}_{j+1} ) 
$$
The deformation space to leading order  is now two dimensional. When we perform a small loop around the singularity in $H$-space, the branch points perform half a turn in opposite directions and at the end of the process they are exchanged. This is the basis of the interpretation by Mich\`ele Audin \cite{Audin01} of the monodromy \cite{Duistermaat80,Zou92,Zung97,Cushman01} using Picard-Lefschetz theory.

\section{Riemann surfaces and integrability.}

In this section we recall, in the example of the Jaynes-Cummings-Gaudin model, the algebro-geometric solution of classical integrable models \cite{DuKrNo90, BaBeTa03, BaTa03, Sk79}.
We first introduce the spectral curve. At each point of the spectral curve we can associate an eigenvector of the Lax matrix. When properly normalised the components of this eigenvector  are meromorphic functions on the spectral curve. The poles of these meromorphic functions are coordinates on phase space. We compute their Poisson brackets. The image of the divisor of the poles by the Abel map is a point on the Jacobian. The motion of this point under the Hamiltonians of the system is linear. We deduce from this that when the spectral curve is non degenerate, the image of the moment map has maximal rank.

\bigskip

The spectral curve is a curve in $\C^2$ defined as $\det ( L(\lambda) - \mu) = 0$ or:
\begin{equation}
\Gamma: \;   \mu^2 - A^2(\lambda) - B(\lambda) C(\lambda) = 0
\label{def_spectral_curve}
\end{equation}
Its importance arises from the fact that it is invariant under the time evolution, so it encodes knowledge
of all the commuting integrals of motion. Equivalently, a spectral curve is attached to each fiber of the moment map. 
In the Jaynes-Cummings-Gaudin model, its equation reads:
$$
\mu^2 = 4\lambda ^2 + 4 H_{n+1}  
+ 2 \sum_{j=1}^{n}  
{H_j  \over \lambda - \epsilon_j}  +  \sum_{j=1}^{n}   {s^2 \over ( \lambda - \epsilon_j)^2 }
$$
We will consider the system reduced by the symmetry generated by $H_{n+1}$. It acts on the Lax matrix by conjugation by a diagonal matrix and obviously leaves the spectral curve invariant. 
Hence the dynamical Hamiltonians are $H_j$, $j=1,\cdots , n$, and $H_{n+1}$ should be considered as a constant parameter.
Note that the dynamical Hamiltonians $H_j$ appear linearly in the equation of the spectral curve:
\begin{equation}
\Gamma : \; R(\lambda,\mu) \equiv R_0(\lambda,\mu) + \sum_{j=1}^g R_j(\lambda) H_j =0
\label{rlambdamu}
\end{equation}
where:
$$
R_j(\lambda) = {2\over \lambda - \epsilon_j},\quad
R_0(\lambda, \mu) = -\mu^2 + 4\lambda^2 + 4 H_{n+1}  + 
\sum_j {s^2 \over ( \lambda - \epsilon_j)^2 }
$$
 As we already noticed
$$
A^2(\lambda) + B(\lambda) C(\lambda) = {Q_{2n+2}(\lambda) \over \prod_j (\lambda - \epsilon_j)^2}
$$
where $Q_{2n+2}(\lambda)$ is a polynomial of degree $2n+2$. Defining $y = \mu  \prod_j (\lambda - \epsilon_j)$,  the equation of the spectral curve becomes
$$
y^2 = Q_{2n+2}(\lambda)
$$
which is an hyperelliptic curve of genus $g= n$. The dimension of the phase space of the model is  $2(n+1)$. However, we  reduced it by the action of the group of conjugation by diagonal matrices which is of dimension 1. Hence we confirm that
$$
g = {1\over 2} {\rm  dim}\;{\cal M}_{\rm reduced} = n
$$

\bigskip

At each point of the  spectral curve, we can solve the equation
$$
\Big( L(\lambda)-\mu\Big) \Psi = 0
$$
Normalizing the second component  (instead of the first one, for later convenience) of $\Psi$ to be $1$, we find:
$$
\Psi = \pmatrix{\psi_1 \cr 1}, \quad \psi_1 = {A(\lambda) + \mu \over C(\lambda)}
$$
showing that $\psi_1$ is a meromorphic functions on the spectral curve. The poles of $\Psi$ at finite distance are located above the zeroes of $C(\lambda)$. Note that 
if $C(\lambda_k) =0$, then the points on $\Gamma$ above $\lambda_k$ have coordinates 
$\mu_k = \pm A(\lambda_k)$. The pole of $\Psi$  is at  the point $ \mu_k =  A(\lambda_k) $, since at the other point, $ \mu_k =  -A(\lambda_k)$, the numerator of $\psi_1$ has a zero. 

\bigskip

 Recalling that 
\begin{equation}
C(\lambda) = 2\bar{b} + \sum_{j=1}^{n} {s_j^+ \over \lambda - \epsilon_j } \equiv
 2\bar{b}\; {\prod_{k=1}^n(\lambda - \lambda_k) \over \prod_{j=1}^{n} (\lambda - \epsilon_j) }
 \label{Csepare}
\end{equation}
we see that indeed the eigenvector has $n=g$ dynamical poles.

\bigskip

At infinity, we have two points:
$$
Q_\pm : \mu = \pm   2 \lambda  (1 + O(\lambda^{-2}) ) 
$$
Remembering that
$$
A(\lambda) =   2 \lambda  + O(\lambda^{-1}), \quad
C(\lambda) = 2\bar{b} + O(\lambda^{-1})
$$
we find the following behavior of the function $\psi_1$ at the two points $Q_\pm$
$$
Q_+ :  \psi_1 = {2\over  \bar{b}}  \; \lambda  + O(\lambda^{-1}),  \quad
Q_-  :  \psi_1 =  O(\lambda^{-1}) 
$$
showing that the eigenvector has a pole at $Q_+$ and a zero at $Q_-$ in agreement with the general
theory. By the Riemann-Roch theorem, the meromorphic function $\psi_1$ exists and is unique.

We can reconstruct the phase-space point of the reduced model (where $H_{n+1}$ is a fixed parameter)
from the knowledge of the $2n$ variables $(\lambda_{k},\mu_{k})$. From eq.~(\ref{Csepare}), we get:
\begin{equation}
s_j^{+}=2\bar{b}\; {\prod_{k=1}^n(\epsilon_{j} - \lambda_k) \over \prod_{i \neq j} (\epsilon_{j} - \epsilon_{i})}
\label{sjz_lambda_mu}
\end{equation}
The $s_{j}^{z}$ components can be obtained from the $n$ equations:
$$ 
\mu_{k}-2\lambda_{k}=\sum_{j=1}^{n}{s_{j}^{z} \over \mu_{k}-\epsilon_{j}} 
$$
after inversion of a Cauchy matrix. 
This analysis shows that the Lax matrix can be reconstructed once we know the coordinates $(\lambda_k,\mu_k)$
of the poles of the eigenvectors. Hence $(\lambda_k,\mu_k)$ can be considered as coordinates on the (reduced) phase space.
We  now compute the symplectic form in these coordinates.
  
From the constraint $(s_j^{z})^2 + s_j^+ s_j^- = s^2$ we can eliminate the variables $s_j^-$.  Remembering the 
Poisson bracket $\{ s_j^z , s_j^+ \} = i s_j^+$, we can write the symplectic form as
$$
\Omega = -i \delta b \wedge \delta \bar{b} +  i  \sum_j  { \delta s_j^+ \over s_j^+}\wedge \delta s_j^z
$$
From eq.(\ref{sjz_lambda_mu}), we get:
$$
{ \delta s_j^+ \over s_j^+} = {\delta \bar{b} \over \bar{b}} + \sum_k {\delta \lambda_k \over \lambda_k - \epsilon_j}
$$
therefore:
$$
\Omega = -i \delta  b \wedge \delta \bar{b} + i  {\delta \bar{b} \over \bar{b}} \wedge \sum_j \delta  s_j^z 
+i \sum_{k} \sum_j {\delta \lambda_k \wedge \delta s_j^z \over \lambda_k - \epsilon_j }
$$
But
$$
\delta  A(\lambda_k) = 2 \delta  \lambda_k + \sum_j { \delta s_j^z \over \lambda_k-\epsilon_j} -
{s_j^z \over (\lambda_k-\epsilon_j)^2 } \delta \lambda_k
$$
and
\begin{eqnarray*}
\Omega &=&  i {\delta \bar{b}\over \bar{b}} \wedge \Big[ \delta (b \bar{b} )+ \sum_j\delta  s_j^z  \Big]
+i \sum_k  \delta \lambda_k \wedge \delta  A(\lambda_k) 
\end{eqnarray*}
Finally:
$$
\Omega = i \delta  \log \bar{b} \; \wedge \delta H_{n+1} + i \sum_k  \delta \lambda_k \wedge \delta  \mu_k
$$
This shows that the variables $(\lambda_k, \mu_k)$ are canonically conjugate. The above calculation is valid 
before the symplectic reduction by $H_{n+1}$.
This expression for the symplectic form confirms that the separated variables are invariant under the diagonal group action
\begin{equation}
\{ H_{n+1},  \lambda_k \} = 0,\quad \{ H_{n+1} , \mu_k \} = 0
\label{Hnlambdak}
\end{equation}
so that they can be used as coordinates on the reduced phase space.

\bigskip

We have found  that $\Gamma$ is of genus $g=n$ and there are exactly $g$ commuting Hamiltonians
$H_j$. Moreover $\Psi$ has exactly $g$ dynamical poles.  The curve is completely determined 
by requiring that it passes through the $g$ points $(\lambda_i, \mu_i)$, $i=1,\cdots, g$. Indeed, the  Hamiltonians $H_j$ are determined by solving the linear system
\begin{equation}
 \sum_j  
{1 \over \lambda_k - \epsilon_j}H_j  = {1\over 2}\mu_k^2 - 2\lambda_k ^2 - 2 H_{n+1}  - {1\over 2} \sum_j {s^2 \over ( \lambda_k - \epsilon_j)^2 }
\label{linsys}
\end{equation}
whose solution is 
\begin{equation}
 H =  B^{-1} V
\label{hamiltoniensclassiques}
\end{equation}
Here the matrix $B_{kj}$ is the Cauchy matrix
\begin{equation}
B_{kj} = {1\over \lambda_k - \epsilon_j}
\label{cauchy}
\end{equation}
and $V_k$ is the right hand side of eq.(\ref{linsys}).

\bigskip

We now compute the equations of motion of the $\lambda_k$'s. One has 
\begin{eqnarray*}
\partial_{t_i} \lambda_k = \{ H_i, \lambda_k \} &=& -\sum_l \{ B_{il}^{-1} V_l , \lambda_k \} =- \sum_lB_{il}^{-1} \{ V_l, \lambda_k\} =- B_{ik}^{-1} \{ V_k, \lambda_k\} 
\label{interesting}
\end{eqnarray*}
where in the last equality, we used the separated structure of the matrix $B$ and the vector $V$ to suppress the sum over $l$.
Explicitely
\begin{equation}
\partial_{t_i} \lambda_k = -i  (B^{-1})_{ik}  \mu_k \quad (\rm{no~summation~over~} k)
\label{dubrovini}
\end{equation}

\bigskip

In order to write the equations of motion  eq.(\ref{dubrovini}) we need to invert the Cauchy matrix. We find
\begin{equation}
(B^{-1})_{jp} = {\prod_{l\neq p} (\epsilon_j -\lambda_l) \prod_i ( \lambda_p-\epsilon_i) \over \prod_{i\neq j} (\epsilon_j-\epsilon_i) \prod_{l\neq p} (\lambda_p-\lambda_l) }
\label{cauchyinverse}
\end{equation}
Hence
\begin{equation}
\partial_{t_i} \lambda_k = -i {\sqrt{Q_{2n+2}(\lambda_k)} \over \prod_{l\neq k} (\lambda_k-\lambda_l)}
{\prod_{l\neq k} (\epsilon_i-\lambda_l) \over \prod_{j\neq i} (\epsilon_i - \epsilon_j) }
\label{floti}
\end{equation}
where the $-i$ comes from the symplectic form.

\bigskip

Note that we can also write equivalently
$$
\sum_k B_{jk}{1\over \mu_k} \partial_{t_i} \lambda_k = -i \delta_{ij}
$$
where
$$
\sigma_j(\lambda_k) = {B_{kj}\over \mu_k}  = {\prod_{l\neq j} (\lambda_k - \epsilon_l) \over \sqrt{Q_{2n+2}(\lambda_k) }} 
$$
but $\sigma_j(\lambda) d\lambda$ are precisely a basis of holomorphic differentials. Hence we  have
\begin{equation}
\sum_k \partial_{t_i} \lambda_k \; \sigma_j(\lambda_k) = -i \delta_{ij}
\label{eqij}
\end{equation}

\bigskip

Define the angles as the images of the divisor $(\lambda_k,\mu_k)$ by the Abel map:
$$
 \theta_j = \sum_k \int^{\lambda_k}  \sigma_j(\lambda,\mu) d\lambda
$$
where $ \sigma_j(\lambda,\mu) d\lambda$ is any basis of holomorphic differentials.
This maps the dynamical divisor $( \lambda_k, \mu_k ), k=1\cdots n$ to a point on the Jacobian of $\Gamma$. We can now prove the fundamental theorem of classical integrable sytems
\begin{theorem}
Under the above map, the flows generated by the Hamiltonians $H_i$ 
are linear on the Jacobian.  
\end{theorem}
\proof
We want to show that the velocities $\partial_{t_i} \theta_j$ are constants, or
$$
 \partial_{t_i} \theta_j = \sum_k \partial_{t_i} \lambda_k \;\sigma_j(\lambda_k,\mu_k) = C^{ste}_{ij}
$$
but this is eq.(\ref{eqij}).
\square

When the divisor $(\lambda_k, \mu_k)$ is in general position, the Jacobi inversion theorem~\cite{Griffiths78} implies that
the matrix $\sigma_j(\lambda_k)$ is invertible and therefore so is the matrix $\partial_{t_i} \lambda_k$. An important consequence of this fact is that the 
projections on the reduced system of the flows $t_i$ are all independent as long as the spectral curve is non degenerate.
In that case, the moment map is of rank $(n+1)$ if the orbit of $H_{n+1}$ is one dimensional, or of rank $n$ if the orbit of $H_{n+1}$ is of dimension zero. But the flow generated by $H_{n+1}$ is just a phase
$$
b(t_{n+1})= e^{it_{n+1}} b(0),\quad s_j^\pm(t_{n+1}) =  e^{\pm it_{n+1}} s_j^\pm(0) 
$$
For the orbit of $H_{n+1}$ to be of dimension zero we must have
$$
b=\bar{b}=0, \quad s_j^\pm = 0, \quad s_j^z = s e_j, \quad e_j=\pm 1
$$
and these are precisely the critical points where the rank of the moment map is zero. Outside these points the rank of the moment map of the full system is equal to the rank of the moment map of the reduced system plus one.

\bigskip

Let us assume that $Q_{2n+2}(\lambda)$ has a double {\em real } root at $\lambda=E$.
$$
Q_{2n+2}(\lambda) = (\lambda - E)^2 \widetilde{Q}_{2n}(\lambda)
$$
 This means that
$A^2(\lambda) + B(\lambda)C(\lambda)$ has a double root. But for real $\lambda$ one has $C(\lambda)= B^*(\lambda)$ so that $A(\lambda)$, $B(\lambda)$, $C(\lambda)$ must all vanish at $\lambda=E$. 

Recalling eq.(\ref{Csepare}), this means\footnote{ If $b(t)= \bar{b}(t) = 0$, then $B(\lambda)$ or $C(\lambda)$ are identicaly zero. The equation of motion $\partial_{t_j} b = ig s_j^-$ implies $s_j^\pm (t)=0$ and therefore $s_j^z(t) = s e_j$. Hence  this corresponds to the singular points.} that one of the separated variables, say $\lambda_1(t)$, is frozen to $E$.
This is compatible with eq.(\ref{floti}) which becomes
\begin{equation}
\partial_{t_i} \lambda_k  = -i {\sqrt{\widetilde{Q}_{2n}(\lambda_k)} \over \prod_{l\neq k,1} (\lambda_k-\lambda_l)}
{\prod_{l\neq k,1} (\epsilon_i-\lambda_l) \over \prod_{j\neq i} (\epsilon_i - \epsilon_j) } (\epsilon_i-E),\quad k=2\cdots n
\label{flotired}
\end{equation}
These flows are not independent however. Because of the identities
$$
\sum_{i=1}^n {\epsilon_i^p \over \prod_{j\neq i} (\epsilon_i-\epsilon_j) } =0,\quad 0\leq p \leq n-2
$$
 we have the relation
$$
\sum_i {1\over \epsilon_i-E} \; \partial_{t_i} \lambda_k =0, \quad k=2\cdots n
$$
On this submanifold the rank of the moment map of the reduced system  is therefore $n-1$, and the moment map of the full system has rank $n$.
By requiring more and more double zeroes and  freezing more and more variables we construct the different strata of the moment map.
This connection between degeneracies of the spectral curve and singularities of the momentum map has been exploited before
for several integrable systems such as spinning tops~\cite{Audin96}.

In the case of a double {\it complex} root $E$, we do have solutions where one $\lambda_{k}$ is frozen to $E$, but these do not exhaust 
all the corresponding fiber of the moment map. A detailed example of this is given in section~\ref{sect_n=1} below, in the case of a system
with one spin ($n=1$).

\bigskip

We arrive at the interesting conclusion that the degeneracies of the moment map and the degeneracies of the spectral curve are intimately related. Recall that the spectral curve reads
\begin{equation}
{Q_{2n+2}(\lambda)\over \prod (\lambda-\epsilon_i)^2  } = 4\lambda^2+  0 \lambda + 4 H_{n+1} +\sum_i {2H_i\over \lambda-\epsilon_i}
+ \sum_i {s^2 \over (\lambda-\epsilon_i)^2}
\label{detL2}
\end{equation}
The allowed real polynomials $Q_{2n+2}(\lambda)$ of degree $2n+2$ are characterized by  the following conditions in the above expansion: 
\begin{eqnarray*}
&&  \textrm{The coefficient of } \lambda^2  \textrm{ is equal to four,} \\
&& \textrm{The coefficient of  }\lambda  \textrm{ is equal to zero,} \\
&&  \textrm{The coefficients of the double poles are } s^2 \Longrightarrow
Q_{2n+2}(\epsilon_i) = s^2 \prod_{j\neq i} (\epsilon_i-\epsilon_j)^2
\end{eqnarray*}

\bigskip

These are $n+2$ conditions on the $2n+3$ {\em real} coefficients of $Q_{2n+2}(\lambda)$. The remaining $n+1$ coefficients are precisely the $n+1$ Hamiltonians $H_i$. The degeneracies we look at are of the form
$$
Q_{2n+2}(\lambda) = \left(\sum_{i=0}^{n+1-r} a_i \lambda^i \right)^2 \left( \sum_{j=0}^{2r} b_j \lambda^j \right)
$$
where the coefficients $a_i, b_j$ are {\em real}. The integer $0 \leq r \leq n+1$ will be the rank of the moment map. To make the decomposition unique we can always impose $a_{n+1-r}=1$ so that we have
$(n+1-r)+2r+1=n+r+2$ coefficients on which we impose  $n+2$ constraints. Hence the leaf of rank $r$ is of dimension $r$.  
We remark that the conditions we have to impose appear as linear equations on the $b_j$ so that we always start by solving them. 
Another consequence of this remark is that if $2r> n+1$ it remains $2r-n-1$ free coefficients $b_j$ which enter the problem linearly. Hence the strata of  rank $r$ contain linear varieties of dimension $2r-n-1$.
In the cases of  rank 0 and rank 1, the genus of the spectral curve is zero.

Strictly speaking, the rank can vary along a given fiber of the moment map. For example, let us
consider a focus-focus critical point in an integrable system with two degrees of freedom. As we have
already mentioned in section~\ref{sec_Normal_Forms} the preimage of the critical value of the moment map is a 
torus which is pinched at the critical point. Therefore, the rank of the moment map for a generic
point on this pinched torus is two, but it falls to zero at the critical point. The above prescription
of freezing $n+1-r$ separated variables $\lambda_k$ on double zeroes of $Q_{2n+2}$   picks configurations
which minimize the rank on the fiber of the moment map defined by the spectral curve.

\section{The one-spin model.}
\label{sect_n=1}
We now study the example of the one-spin system from the point of view of the degeneracies of the spectral curve.  This model is very well known in the physical literature \cite{JC} but  also appeared recently in the mathematical literature \cite{San10}.
In the one-spin case, the Hamiltonians read
\begin{eqnarray*}
H_1&=& 2\epsilon_1  s_1^z + b s_1^+ + \bar{b}s_1^-   \\
H_2 &=& \bar{b} b + s_1^z 
\end{eqnarray*}

Recall again the spectral curve eq.(\ref{detL}) which reads in this case :
\begin{equation}
{Q_4(\lambda)\over (\lambda-\epsilon_1)^2 } = 4\lambda^2+ 4 H_2 + 2{H_1\over \lambda-\epsilon_1}
+ {s^2 \over (\lambda-\epsilon_1)^2}
\label{detL1}
\end{equation}
We want to see when the spectral curve degenerates. 

\subsection{Rank 0:}
As we have seen the singular points are given by  $b=\bar{b} = 0, s^{\pm}_1 = 0$. Hence we have two points 
$$
s_1^z = e s, \quad e=\pm 1
$$
The corresponding values $P=(H_1,H_2)$ are
$$
P_1(\uparrow)=\left[ 2\epsilon_1 s, s \right] 
$$
$$
P_2(\downarrow)=\left[ -2\epsilon_1 s, -s \right] 
$$

We can recover this result by analysing the spectral curve. The most degenerate case is when $Q_4(\lambda)$ is a perfect square.
Assuming 
$$
Q_4(\lambda)= (a_2 \lambda^2 + a_1 \lambda + a_0)^2, \quad a_2 \neq 0
$$
We have three coefficients $a_i$ on which we impose three conditions hence they are completely determined. We find
$$
a_2 \lambda^2 + a_1 \lambda + a_0 = 2\left( \lambda^2 -\epsilon_1 \lambda + {s\over 2} e_1 \right)
$$
Comparing the partial fraction decomposition of both sides of eq.(\ref{detL1}),  we find
the corresponding values of $(H_1,H_2)$. They are precisely $P_1$ and $P_2$. Hence the points of rank zero are the only points where the spectral curve is totally degenerate.

\bigskip

To determine the type of the singularities, we look at the classical Bethe equations which  read 
$$
2\mu + {se\over \mu-\epsilon_1}=0 \Leftrightarrow 2\mu^2 -2\epsilon_1 \mu + se =0
$$
The discriminant of this equation is  $\epsilon_1^2 - 2 s e $. So, when the spin is down (point $P_2$), the discriminant is positive,
the two classical Bethe roots are real and this is an elliptic singularity, in agreement with the general analysis of section
 \ref{sec_Normal_Forms}. When the spin is up (point $P_1$) we have real roots when $\epsilon_1^2 \geq 2s$ (i.e. the singularity is elliptic in this case), and  a pair of complex conjugate roots $E,\bar{E}$ 
when $\epsilon_1^2 \leq 2s$ (i.e. the singularity is focus-focus in that case).

\begin{figure}[ht]
\begin{center}
\includegraphics[height= 4cm]{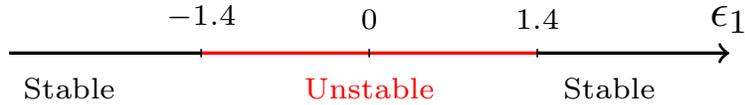}
\caption{ The domains of  the coupling constant $\epsilon_1$, $(s=1)$.}
\label{domain1spin}
\end{center}
\nonumber
\end{figure}

Let us now discuss the fibers of the moment map over the critical values $P_1$ and $P_2$. In the stable case,
such fiber is reduced to the critical point. But in the unstable case (focus-focus singularity), this fiber
is a two-dimensional torus pinched at the critical point. After the symplectic reduction associated to $H_2$,
the pinched torus is conveniently described as a finite arc in the complex plane of the separated variable $\lambda_1$.  
The spectral polynomial can be written as:

\begin{equation}
Q_4(\lambda) = 4(\lambda -E)^2(\lambda-\bar{E})^2 
\label{degen1}
\end{equation}
where:
$$
\epsilon_1=E+\bar{E}, \quad E\bar{E} = {s\over 2}
$$ 
The separated variable $\lambda_1$ is defined by:
$$
C(\lambda) = 2\bar{b} {\lambda-\lambda_1\over \lambda -\epsilon_1} = 2\bar{b} + {s_1^+\over \lambda-\epsilon_1} 
$$
so $s_1^{+}$ is expressed as:
\begin{equation}
s_1^+=2\bar{b}(\epsilon_1-\lambda_1)
\label{expr_s1+}
\end{equation}
The conjugated variable $\mu_1$ is equal to
$$
\mu_1=A(\lambda_1)=2\lambda_1+{s_{1}^{z} \over \lambda-\epsilon_1} 
$$
so that:
\begin{equation}
s_1^z = (\epsilon_1-\lambda_1) (2\lambda_1 - \mu_1)
\label{s+szbis}
\end{equation}
Since $(\lambda_1,\mu_1)$ belongs to the spectral curve, we have:
$$
\mu_1 = \pm 2 {(\lambda_1-E)(\lambda_1-\bar{E}) \over (\lambda_1-\epsilon_1)}
$$
Choosing the plus sign leads to $s_1^z = s$ which is the unstable point. So we now
choose the minus sign, which gives:
\begin{equation}
s_1^z =-2\left(\lambda_1(\lambda_1-E-\bar{E})  + (\lambda_1-E)(\lambda_1-\bar{E})\right)
\label{s+szter}
\end{equation}
The equation of motion for the flow generated by $H_1$ is:
\begin{equation}
\partial_{t_1} \lambda_1 = 2i (\lambda_1 -E)(\lambda_1-\bar{E})
\label{flot1spin}
\end{equation}
whose solution is: 
$$
\lambda_1 = {E-\bar{E}X \over 1 - X}, \quad X= A_1 e^{2i(E-\bar{E})t_1} 
$$
From the expression of $s_{1}^{+}$ eq.~(\ref{expr_s1+})
and the equation of motion
\begin{equation}
\partial_{t_1}\bar{b} =- i s_1^+
\label{mouvb}
\end{equation}
we deduce:
$$
\bar{b}(t) = \bar{b}_0 {e^{-2i\bar{E} t_1} \over 1-X }, \quad s_1^+ = 2\bar{b}_0\;\;\;
{ \bar{E} -E X  \over (1-X )^2 } \; e^{-2i\bar{E}t_1} 
$$
From eq.(\ref{s+szter}) we get:
$$
s_1^z = 2 E\bar{E} -4(E-\bar{E})^2 {X\over (1-X)^2}
$$
Now $s_1^z$ should be real, which is equivalent to $(X-\bar{X})(X\bar{X}-1)=0$. With $X= A_1 e^{2i(E-\bar{E})t_1}$ it is impossible to have
$X\bar{X}=1$ for all $t_1$. Then $X=\bar{X}$, which imposes $\bar{A_1} = A_1$ so that $A_1$ is real. Its absolute value can be absorbed  
in the origin of time $t_1$. Only its sign matters. To determine this sign, let us impose the constraint that the length of the spin is constant. 
This gives the condition:
$$
\bar{b}_0  b_0 = 4 A_1(E-\bar{E})^2
$$
Then:
$$
(s^z)^2 + s^+s^- = 4 E^2\bar{E}^2 = s^2,
$$
We see that $A_1$ should be negative. As a consequence, $\lambda_1$ runs along the line interval joining $E$ and $\bar{E}$. Here, we see
explicitely that freezing $\lambda_1$ on $E$ (or on $\bar{E}$) gives only the critical point $(\uparrow)$ which is only a tiny part
of the fiber above the critical value $P_{1}$ in the focus-focus case. Note that on this fiber the rank of the moment map is two,
excepted fot the critical point $(\uparrow)$ where it vanishes.

\bigskip

\subsection{Rank 1} We assume next that
\begin{equation}
Q_4(\lambda) = (\lambda + a_0)^2 (b_2 \lambda^2 + b_1 \lambda + b_0), \quad b_2 \neq 0
\label{Q42}
\end{equation}
where $a_0$ and $b_i$ are real. We denote:
$$
a_0={1\over 2} \;x - \epsilon_1 
$$
Imposing the three conditions on $Q_4(\lambda)$, we can determine $b_0, b_1,b_2$ in term of $x$ by solving linear equations. We find:
$$
b_2= 4, \quad
b_1= -4 x, \quad
b_0= 4\;{  \epsilon_1 x^3 - \epsilon_1^2 x^2 +  s^2 \over  x^2 }
$$
\begin{eqnarray*}
H_1 &=& -{x^4 -2\epsilon_{1} x^3 -4 s^2 \over 2 x} \\
H_2 &=& -{3 x^4 -8\epsilon_{1} x^3 + 4 \epsilon_{1}^2 x^2 -4 s^2 \over 4 x^2}
\end{eqnarray*}
and we see that: 
\begin{equation}
{\partial H_1\over \partial x}= x{\partial H_2\over \partial x}
\label{see_rank1_n=1}
\end{equation}
Since we are on a rank one line, the derivatives of the functions $H_1, H_2$  with respect to any coordinates $X$ and $Y$ on phase space, evaluated on the line are proportional:
$$
{\partial \over \partial X} \pmatrix{H_1\cr H_2} \propto {\partial \over \partial Y} \pmatrix{H_1\cr H_2} \Longrightarrow 
{\partial_X H_1\over \partial_X H_2}={\partial_Y H_1\over \partial_Y H_2} =x
$$
In particular  we have:
$$
{\partial  H_1\over \partial b}  = x \; {\partial  H_2\over \partial b} \Longrightarrow s_1^+ =x\bar{b}
$$

%From the previous analysis, the divisor is trapped at $\lambda_1=-a_0= \epsilon_1-{x\over 2}$. Then 
%$$
%s_1^+ = 2\bar{b}(\epsilon_1-\lambda_1)=x\bar{b}
%$$

Inserting this relation into the definitions of $H_1$ and $H_2$ gives: $H_2= \bar{b}b+s_1^z$ and $H_1=2\epsilon_1 s_1^z + 2x\bar{b}b$. Therefore:
$$
s_1^z = -{1\over 2} x(x-2\epsilon_1)
$$
$$
\bar{b}b ={ (2s -x^2 +2\epsilon_1 x )(2s +x^2 -2\epsilon_1 x )\over 4 x^2}
$$
Notice that when 
\begin{equation}
x^2 -2\epsilon_1 x + 2 e s = 0, \quad e=\pm 1
\label{xbound}
\end{equation}
we have $s_1^z = es$ and $\bar{b} b=0$ and this corresponds to the points $P_1$ and $P_2$.  In fact there is a simple relation between rank 0 and rank 1. 
Rank 1 spectral curves degenerate when the polynomial $b_2 \lambda^2+ b_1 \lambda + b_0$ has a double real root. Its discriminant is:
$$
b_1^2-4 b_0 b_2 = 16\; { (x^2 -2\epsilon_1 x -2s)(x^2 -2\epsilon_1 x +2s)\over x^2}=-64\; \bar{b} b \leq 0
$$
It vanishes precisely when eqs.(\ref{xbound}) are satisfied. 
These equations are nothing but the classical Bethe equations eq.(\ref{classicalBethe}) as can be seen by setting
$$
x=2 \mu
$$
Remark that the boundary of the image of the moment map is obtained for $x$ real, because $H_1$ and $H_2$ are both real. 
Hence the elliptic points are on the boundary, but the focus-focus points correspond to complex $x$ and therefore are not
on this boundary (see fig.~\ref{instablepolytope} below).
To end the characterization of the boundary of the moment map we have to determine the range of $x$. 
The physical constraints are $-s \leq s_1^z \leq s$ and $\bar{b}b \geq 0$. These two conditions are both
equivalent to:
\begin{equation}
-2s +\epsilon_1^2 \leq (x-\epsilon_1)^2 \leq 2s+\epsilon_1^2
\label{physical_condition_n=1}
\end{equation}
In this inequality we recognize the discriminants $\epsilon_1^2\pm 2s$
of the classical Bethe equations, so that we have to distinguish between the stable and unstable case.

\subsubsection{Stable case.}
\begin{figure}[h!]
\begin{center}
\includegraphics[height= 4cm]{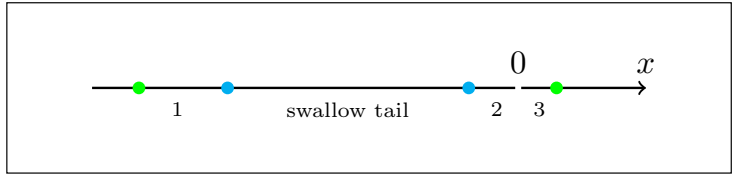}
\caption{The range of $x$. The allowed segments are labelled $1$, $2$, $3$. The points of the same color are mapped to the same singular point in the bifurcation diagram. Notice that $x=0$ is mapped to infinity.  $(\epsilon_1=-2)$.}
\label{rangestable}
\end{center}
\nonumber
\end{figure}

\begin{figure}[h!]
\begin{center}
\includegraphics[height= 8cm]{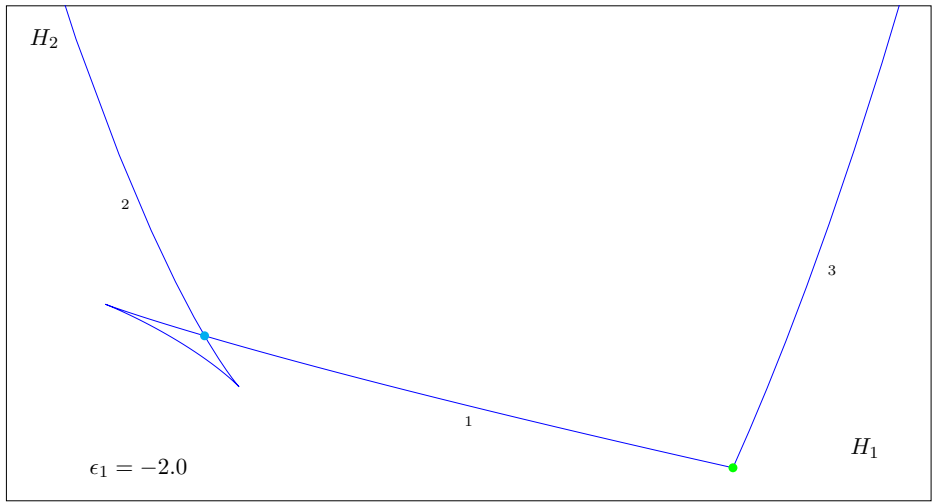}
\caption{Image of the moment map in the stable case. The swallow tail is not part of the image of the moment map. $(\epsilon_1=-2)$.}
\label{stablepolytope}
\end{center}
\nonumber
\end{figure}

When $\epsilon_1^2 \geq 2s$ the two sides of the inequality are positive and put bounds on $x$. The two equations determining the boundary values of $x$ are precisely eqs.(\ref{xbound}), and we are in the case where all their roots are real. The allowed range of $x$ is as in fig.(\ref{rangestable}). From this we can construct the image of the moment map. It is shown in fig.(\ref{stablepolytope}) with the edges labelled according to the ranges of $x$. The point $P_2$ (spin down) is the green point and the point $P_1$ (spin up) is the cyan point. They are both on the boundary of the image.

The swallow tail is not part of the image of the moment map. It is composed of the points $(H_{1}(x),H_{2}(x))$ 
when $x$ runs in the interval $[\epsilon_{1}-\sqrt{-2s+\epsilon_{1}^{2}},\epsilon_{1}+\sqrt{-2s+\epsilon_{1}^{2}}]$.
The two cusps correspond to values $x_{c}$ of $x$ such that $\partial H_1/\partial x$ and $\partial H_2/\partial x$
both vanish. Let us write $x=x_{c}+u$. Then, taking into account eq.~(\ref{see_rank1_n=1}), we have the following Taylor expansions
for small u:
\begin{eqnarray}
\delta H_{2} & = & {\alpha \over 2}u^{2}+{\beta \over 3}u^{3}+\mathcal{O}(u^{4}) \label{delta_H1}\\
\delta H_{1} & = & x_{c} \delta H_{2}+{\alpha \over 3}u^{3}+\mathcal{O}(u^{4})
\end{eqnarray}
where $\delta H_{i}$ stands for the the small variation $H_{i}(x_{c}+u)-H_{i}(x_{c})$.
From eq.~(\ref{delta_H1}), we have:
$$
u=\pm\left({2\over \alpha} |\delta H_{2}| \right)^{1/2}+\mathcal{O}(\delta H_{2})
$$
Therefore:
\begin{equation}
\delta H_{1}=x_{c}\delta H_{2}\pm {\alpha \over 3}\left({2\over \alpha} |\delta H_{2}| \right)^{3/2}+\mathcal{O}((\delta H_{2})^{2})
\end{equation}
which gives the leading shape of the cusp near $x_{c}$. We see that the two branches are found on opposite sides of their common tangent,
as shown on fig.~\ref{stablepolytope}.

\subsubsection{Unstable case.}
\begin{figure}[h!]
\begin{center}
\includegraphics[height= 4cm]{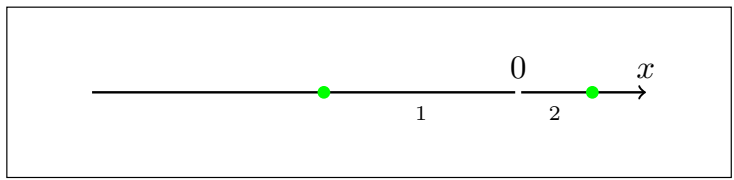}
\caption{The range of $x$. The green points correspond to the stable point $P_2$ and are solutions 
of the equation $x^2-2\epsilon_1 x -2s =0$.
Since points of the same color are mapped to the same point by the moment map, and zero is mapped to $\infty$,  
the boundaries of the image of this map are easy to reconstruct.  $(\epsilon_1= -0.707)$.}
\label{rangeinstable}
\end{center}
\nonumber
\end{figure}

\begin{figure}[h!]
\begin{center}
\includegraphics[height= 8cm]{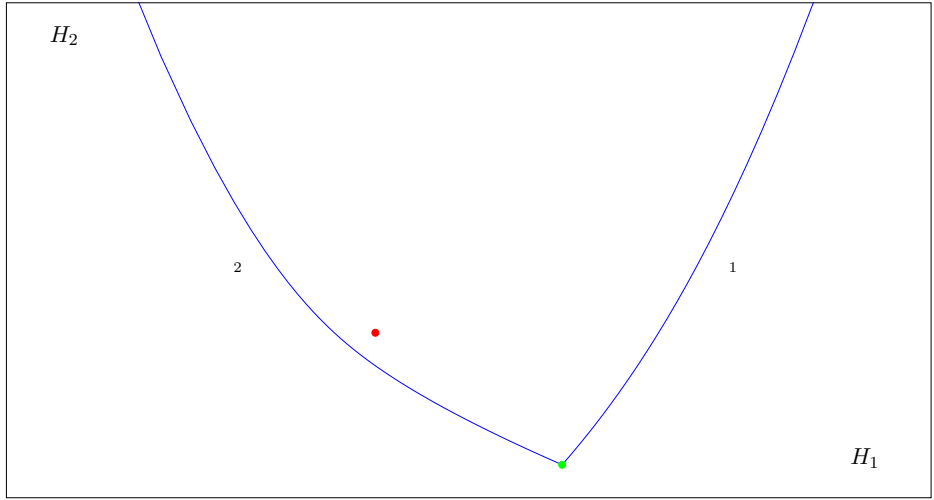}
\caption{The image of the moment map in the case of one spin, with one unstable critical point. 
The green point is the stable (elliptic) point $P_2$. The red point is the unstable (focus-focus) point $P_1$.  
It is in the interior of the image of the moment map.
$(\epsilon_1=-0.707)$.}
\label{instablepolytope}
\end{center}
\nonumber
\end{figure}

When $\epsilon_1^2 \leq 2s$ the left side of the inequality~(\ref{physical_condition_n=1}) 
is always satisfied. Among the equations eqs.(\ref{xbound}) determining the bounds of $x$, 
one has two real roots and the other one has two complex conjugate roots. 
The range of $x$ is as in fig.(\ref{rangeinstable}). The image of the moment map is shown in fig.(\ref{instablepolytope}).  
The point $P_2$ (spin down) is the green point and is a vertex on the boundary. 
The point $P_1$ (spin up) is the red point. It is in the interior of the image of the moment map.

\newpage

\section{The two-spins model.}

For the two spins model, let us write explicitly the Hamiltonians:
\begin{eqnarray*}
H_1&=& 2\epsilon_1 s_1^z + b s_1^+ + \bar{b}s_1^-  + {s_1\cdot s_2 \over \epsilon_1-\epsilon_2} \\
H_2&=& 2\epsilon_2  s_2^z + b s_2^+ + \bar{b}s_2^-   -  {s_1\cdot s_2 \over \epsilon_1-\epsilon_2} \\
H_3 &=& \bar{b} b + s_1^z + s_2^z
\end{eqnarray*}
where
$$
s_1\cdot s_2={s_1^z}\,{s_2^z}+{1\over{2}} ({{s_1^-}\,{s_2^+}+{s_1^+}\,{s_2^-}})
$$

The spectral curve reads in this case:
\begin{equation}
{Q_6(\lambda)\over (\lambda-\epsilon_1)^2 (\lambda-\epsilon_2)^2 } = 4\lambda^2+ 4 H_3 + {2H_1\over \lambda-\epsilon_1}
+ {2H_2\over \lambda-\epsilon_2}
+ {s^2 \over (\lambda-\epsilon_1)^2}+ {s^2 \over (\lambda-\epsilon_2)^2}
\label{detL2_n=1}
\end{equation}

Let us now use the degeneracies of zeroes of $Q_6(\lambda)$ to study the rank of the moment map.

\subsection{Rank 0} 

Again, the singular points are given by  $b=\bar{b} = 0, s^{\pm}_1 = s^{\pm}_2 =0$ so that we have four critical points: 
$$
s_1^z = \pm s,\quad s_2^z = \pm s
$$

The corresponding values $P=(H_1,H_2,H_3)$ are:
\begin{eqnarray}
P_1(\uparrow,\uparrow)&=&\left[ {{s^2}\over{\epsilon_1-\epsilon_2}}+2\,\epsilon_1\,s ,\quad 2\,\epsilon_2\,s-{{s^2
 }\over{\epsilon_1-\epsilon_2}} ,\quad 2\,s \right]  \label{defP1} \\
P_2(\uparrow,\downarrow)&=&\left[ 2\,\epsilon_1\,s-{{s^2}\over{
 \epsilon_1-\epsilon_2}} , \quad {{s^2}\over{\epsilon_1-\epsilon_2}}
 -2\,\epsilon_2\,s ,\quad 0 \right]  \label{defP2} \\
P_3(\downarrow,\uparrow)&=&\left[ -2\,\epsilon_1\,s-{{s^2}\over{
 \epsilon_1-\epsilon_2}} ,\quad  {{s^2}\over{\epsilon_1-\epsilon_2}}+
 2\,\epsilon_2\,s ,\quad 0 \right]  \label{defP3} \\
 P_4(\downarrow,\downarrow)&=&\left[ {{s^2}\over{\epsilon_1-\epsilon_2}}
 -2\, \epsilon_1\,s , \quad -2\,\epsilon_2\,s-{{s
 ^2}\over{\epsilon_1-\epsilon_2}} ,\quad -2\,s \right] \label{defP4}
 \end{eqnarray}

We can recover this result by analysing the degeneracies of the spectral curve. The most degenerate case corresponds to a polynomial $Q_6(\lambda)$ of the form
$$
Q_6(\lambda) = (a_3\lambda^3+ a_2\lambda^2+a_1\lambda + a_0)^2
$$
The four conditions we have to impose on $Q_6(\lambda)$ determine completely the four coefficients $a_i$. We find:
\begin{equation}
a_3\lambda^3+ a_2\lambda^2+a_1\lambda + a_0 = 2
\left(\lambda^3 -(\epsilon_1+\epsilon_2) \lambda^2 
+\left( \epsilon_1\epsilon_2   + {s\over 2}(e_1+e_2) \right)  \lambda -{s\over 2} (e_1\epsilon_2+e_2\epsilon_1) \right)
\label{Q6R0}
\end{equation}
where $e_i=\pm 1$. The values of the energies are exactly eqs.(\ref{defP1}--\ref{defP4}).

In order to determine the type of the singularities we  write the classical Bethe equations:
 \begin{equation}
2 \mu +   { s e_1\over \mu-\epsilon_1} + {s e_2\over \mu-\epsilon_2} =0
 \label{eqE1}
 \end{equation}
 These are polynomial equations of degree 3. The number of real roots is determined by the sign of the discriminant:
 \begin{eqnarray*}
{\rm Disc}(e_1,e_2)&=& -4\,\epsilon_1^2 \epsilon_2^2\left(\epsilon_2-\epsilon_1\right)^2 
+4\left(\epsilon_2-\epsilon_1\right)\left(2\,{ e_1}
 \epsilon_2^3+{ e_2}\,\epsilon_1\,\epsilon_2^2-2{ e_1}
 \epsilon_1\epsilon_2^2+2{ e_2}\epsilon_1^2\epsilon_2-
 { e_1}\epsilon_1^2\epsilon_2-2 e_2\epsilon_1^3
 \right)s  \\
&& -\left(\epsilon_2^2+20 e_1 e_2\epsilon_2^2-8\epsilon_2^2+8\epsilon_1 \epsilon_2
-38e_1e_2 \epsilon_1\epsilon_2+8\epsilon_1\epsilon_2-8\,\epsilon_1^2+20  e_1 e_2\epsilon_1^2+\epsilon_1^2 \right)s^2  \\
&& +2\left(e_2+e_1\right)^3 s^3 
 \end{eqnarray*}
 
\begin{figure}[h!]
\begin{center}
\includegraphics[height= 10cm]{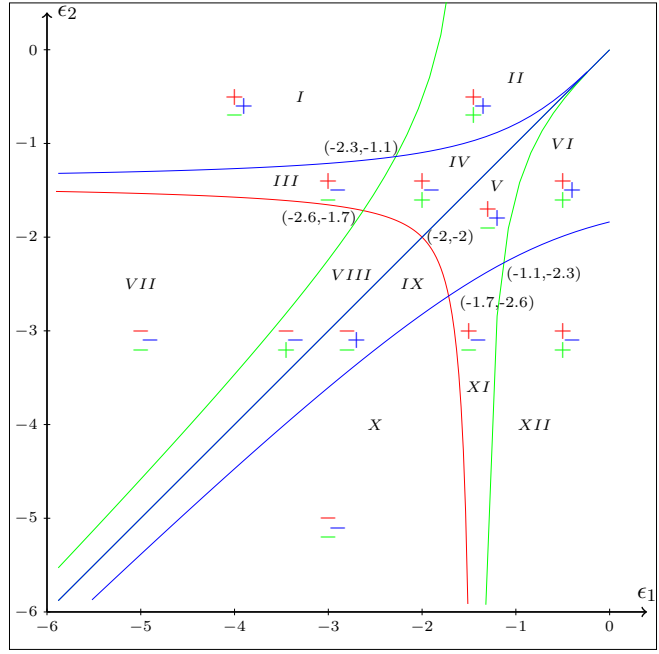}
\caption{The domains indicating the signs of the various discriminants, for three critical points of the system
with two spins. Curves correspond to the vanishing of one discriminant. 
The color of these curves and of signs inside the domains label the critical
points according to the following code: red stands for up-up, green for up-down, and blue for down-up. $(s=1)$.}
\label{domainepsilon}
\end{center}
\nonumber
\end{figure}

 If ${\rm Disc}(e_1,e_2) <0$ we have three real roots (elliptic point) and if ${\rm Disc}>0$ we have one real root
(focus-focus singularity). 
 We restrict ourselves to the case where $\epsilon_1$ and $\epsilon_2$ are both negative. We easily see that ${\rm Disc}(e_1=-1,e_2=-1)$ is always negative, hence the configuration with two spins down is always stable. The signs of the other three discriminants is shown in fig.(\ref{domainepsilon}).

 In region $V$, for instance, we have two unstable points and two stable points. 
 \begin{eqnarray*}
(\uparrow,\uparrow)&:& {\rm one~real~root~~(unstable)} \\
(\uparrow,\downarrow)&:& {\rm three~real~roots~~(stable)} \\
(\downarrow,\uparrow)&:& {\rm one~real~root~~(unstable)} \\
(\downarrow,\downarrow)&:& {\rm three~real~roots~~(always~stable)}
\end{eqnarray*}

\subsection{Rank 1} 
 
 We now set
\begin{equation}
Q_6(\lambda) = (\lambda^2 + a_1\lambda + a_0)^2 (b_2\lambda^2+b_1\lambda+b_0)
\label{Q6R1}
\end{equation}
We have five coefficients and four conditions on $Q_6$. Hence we have a dimension one manifold of solutions. The coefficients $b_j$ are completely determined and there is one constraint between $(a_0,a_1)$.
Note that the case of rank 0 is obtained as a special case of rank 1, when the polynomial
$b_2\lambda^2+b_1\lambda+b_0$ has a doubly degenerate root.
Let us parametrize $(a_0,a_1)$ in terms of $(x,y)$ as follows:
$$
a_0=-{{\epsilon_1\,y+\epsilon_2\,x-2\,\epsilon_1\,\epsilon_2 }\over{2}}, \quad 
a_1={{y+ x-2\,\epsilon_2-2\,\epsilon_1}\over{2}}
$$
Imposing the coefficient of $\lambda^2$ we find:
$$
b_2=4
$$
Imposing the vanishing of the coefficient of $\lambda$, we find:
$$
b_1=-4\,\left(y+x\right)
$$
Imposing that the coefficient of the double pole at $\lambda=\epsilon_2$ is $s^2$ we find:
$$
b_0={{4\,\left(\epsilon_2\,y^3+\epsilon_2 \,x\,y^2-\epsilon_2^2\,y
 ^2+s^2\right)}\over{y^2}}
 $$
Imposing next that the coefficient of the double pole at $\lambda=\epsilon_1$ is $s^2$ we find the constraint: 
 \begin{equation}
 S_1:\quad   s^2( y^2 - x^2 )+ (\epsilon_1-\epsilon_2) (x+y-\epsilon_1-\epsilon_2)x^2 y^2=0
 \label{S1}
 \end{equation}
Then $H_1, H_2, H_3$ are given by eqs.(\ref{H1XY},\ref{H2XY},\ref{H3XY}).

 \begin{eqnarray}
 H_1&=& {1\over{2\,\left( \epsilon_2-\epsilon_1\right)\,y^2}}
 \left( {\left(\epsilon_2-\epsilon_1\right)
  \left( -2\,\,x\,y^4-3\,\,x^2\,y^3-2\,\left(-3\,\epsilon_2+\epsilon_1\right)\,
 x\,y^3 -\,x^3\,y^2 \right) } \right.  \nonumber\\
&&
\left. -\left( \epsilon_2-\epsilon_1\right)\,\left(-4\,\epsilon_2+2\,
 \epsilon_1\right)\,x^2\,y^2+4\,\left(\epsilon_2-\epsilon_1\right)^2
 \,\left(-\epsilon_2-\epsilon_1\right)\,x\,y^2-2\,s^2\,x\,
 y \right. \nonumber \\
 && \left.   +4\,\left(\epsilon_2-\epsilon_1\right)\,s^2\,x \right)    \label{H1XY}
 \end{eqnarray}
 
 \begin{equation}
 H_2={{-\left(\epsilon_2-\epsilon_1\right)\,y^4-\left(\epsilon_2-
 \epsilon_1\right)\,x\,y^3-\left(\epsilon_2-\epsilon_1\right)\,
 \left(-2\,\epsilon_2\right)\,y^3+2\,s^2\,x+4\,\epsilon_2
 \,s^2-4\,\epsilon_1\,s^2}\over{2\,\left(\epsilon_2-\epsilon_1
 \right)\,y}} \label{H2XY}
 \end{equation}
 
 \begin{equation}
 H_3={{-3\,y^4-6\,x\,y^3-4\,\left(-2\,\epsilon_2
 \right)\,y^3-3\,x^2\,y^2-4\,\left(-\epsilon_2-
 \epsilon_1\right)\,x\,y^2-\left(-2\,\epsilon_2\right)^2\,y^2+4
 \,s^2}\over{4\,y^2}} \label{H3XY}
 \end{equation}
These are the parametric equations of the lines of rank 1. The parameters $x$ and $y$ are tied together by relation eq.(\ref{S1}).
 
Let us now define the total derivative with respect to $x$ by: 
 $$
 {d\over dx} = {\partial \over \partial x} + {d y \over d x}\; {\partial \over \partial y}, \quad \mathrm{where}\;\; 
 {d y \over d x} = -{\partial _x S_1\over \partial_y S_1}
 $$
Then we can compute: 
\begin{eqnarray*}
{d H_1 \over d x} - x {d H_3 \over dx} &=& -{ y + 2(\epsilon_1-\epsilon_2) \over (\epsilon_1 -\epsilon_2) x y^3 }  {d y \over d x} \; S_1 \\
{d H_2 \over d x} - y {d H_3 \over dx} &=& { 2 \over (\epsilon_1 -\epsilon_2) x y^2 }  {d y \over d x} \; S_1
\end{eqnarray*}
hence, on $S_1$, we have: 
$$
{dH_1\over dx}=x \; {d H_3\over dx} ,\quad {dH_2\over dx}=y \; {d H_3\over dx} 
$$
Because we are on a rank one line this relation is true for any derivative on the line.
Considering the derivative with respect to $b$ we find
$$
s_1^+ = x \bar{b}, \quad s_2^+ = y \bar{b}
$$
and considering the derivative with respect to $s_1^+$ we get:
$$
2 s_1^z +{1\over \epsilon_1-\epsilon_2} (-x s_2^z + y s_1^z ) = -x^2+2\epsilon_1 x ,\quad {1\over \epsilon_1-\epsilon_2} (-x s_2^z + y s_1^z ) = x y
$$
from which we deduce:
$$
s_1^z = -{x\over 2} (x+y-2\epsilon_1),\quad s_2^z = -{y\over 2} (x+y-2\epsilon_2)
$$
and:
$$
\bar{b} b = -{(y^2+xy -2\epsilon_2 y -2s)(y^2+xy -2\epsilon_2 y +2s) \over 4 y^2}
$$

Note that the discriminants of the two second degree polynomials which appear in the factorization (\ref{Q6R1}) of 
$Q_{6}(\lambda)$ are:

\begin{eqnarray*}
b_1^2-4 b_0 b_2 &=& {16(y(x+y)-2\epsilon_2 y - 2s ) (y(x+y)-2\epsilon_2 y + 2s )\over  y^2 } = -64 \bar{b} b <0 \\
a_1^2-4a_0 &=& {1\over 4} \Big( (x+y)^2 -4(\epsilon_1-\epsilon_2)(x-y) + 4(\epsilon_1-\epsilon_2)^2\Big) \equiv {1\over 4} \Delta
\end{eqnarray*}
%where
%$$
%\Delta = (x+y)^2 -4(\epsilon_1-\epsilon_2)(x-y) + 4(\epsilon_1-\epsilon_2)^2
%$$
The first discriminant vanishes when $Q_{6}(\lambda)$ has three double roots, that is when the rank
drops to zero. In this case, inserting the parametrization
$$
x=-s {e_1\over \mu-\epsilon_1},\quad y=-s{e_2\over \mu-\epsilon_2}
$$
into eq.(\ref{S1}), we see that it turns into the classical Bethe equation for $\mu$.
The variables $x,y$ being real by definition, the real solutions of the classical Bethe equations correspond to points on the curve $S_1$.

When $\Delta>0$, $Q_{6}(\lambda)$ has two real double roots, which become a pair of complex conjugated double roots
when $\Delta<0$.  Most likely, the sign of $\Delta$ governs the nature of the preimage of points lying on such a line
of rank 1. To check this, one would need to compute the corresponding normal forms, which we have not done yet, the normal
forms discussed previously being specialized to the special case of rank zero singular points. We leave this generalization
as a subject for future work. We expect that points where $\Delta=0$ correspond to qualitative changes in the topology of the
pre-image. They are determined by the following factorization for $Q_{6}(\lambda)$: 
$$
Q_6(\lambda)=4(\lambda-a_0)^4(\lambda^2+b_1\lambda+b_0)
$$
As before, we express the known constraints on this polynomial, which gives:
$$
4\,\left(-a_0+\epsilon_1\right)^4\,\left(b_0+b_1\,\epsilon_1+
 \epsilon_1^2\right)-\left(\epsilon_1-\epsilon_2\right)^2\,s^2=0
$$ 
$$
4\,\left(-a_0+\epsilon_2\right)^4\,\left(b_0+b_1\,\epsilon_2+
 \epsilon_2^2\right)-\left(\epsilon_1-\epsilon_2\right)^2\,s^2=0
$$ 
$$
-16\,a_0+4\,b_1+8\,\epsilon_1+8\,\epsilon_2=0
$$ 
Solving for $b_0$ and $b_1$ in the first two equations gives: 
$$
b_0=\epsilon_1\,\epsilon_2-{{\left(-\epsilon_1^2+\epsilon_1\,\epsilon_2\right)\,s^2}
\over{4\,\left(a_0-\epsilon_2\right)^4}}+{{\left(-\epsilon_1\,\epsilon_2+\epsilon_2^2\right)\,s^2}
\over{4\,\left(a_0-\epsilon_1\right)^4}}
$$
and 
$$
b_1=-\epsilon_1-\epsilon_2-{{\left(-\epsilon_1+\epsilon_2\right)\,s^2}
\over{4\,\left(a_0-\epsilon_1\right)^4}}+{{\left(-\epsilon_1+\epsilon_2\right)\,s^2}
\over{4\,\left(a_0-\epsilon_2\right)^4}}
$$ 
Plugging these values in the third equation determines $a_0$ through: 
$$
-16\,a_0+4\,\epsilon_1+4\,\epsilon_2+{{\left(\epsilon_1-\epsilon_2\right)\,s^2}
\over{\left(a_0-\epsilon_1\right)^4}}+{{\left(-\epsilon_1+\epsilon_2\right)\,s^2}
\over{\left(a_0-\epsilon_2\right)^4}}=0
$$

\subsection{Rank 2}  

We now set:
\begin{equation}
Q_6(\lambda) = (\lambda  + a_0)^2 (b_4\lambda^4+b_3\lambda^3 +b_2\lambda^2+b_1\lambda+b_0)
\label{Q6R2}
\end{equation}

The four constraints are linear equations determining $b_4,b_3,b_2,b_1$. It remains two parameters $a_0,b_0$. 
The rank 2 manifolds are two-dimensional surfaces. Remark that $b_0$ appears linearly. The four conditions on $Q_6$ read:
\begin{eqnarray*}
b_4 & = & 4 \\
\left(\epsilon_1+a_0\right)^2\,\left(b_4\,\epsilon_1^4+
b_3\,\epsilon_1^3+b_2\,\epsilon_1^2+b_1\,\epsilon_1+b_0\right) & = & \left(\epsilon_1-\epsilon_2\right)^2\,s^2 \\
\left(\epsilon_2+a_0\right)^2\,\left(b_4\,\epsilon_2^4+
b_3\,\epsilon_2^3+b_2\,\epsilon_2^2+b_1\,\epsilon_2+b_0\right) & = & \left(\epsilon_1-\epsilon_2\right)^2\,s^2 \\ 
2\,b_4\,\epsilon_2+2\,b_4\,\epsilon_1+2\,a_0\,b_4+b_3 & = & 0 
\end{eqnarray*}
We can solve them in terms of $a_0$ and $b_0$:
\begin{eqnarray*}
b_4 & = & 4 \\
b_3 & = & -8\,\left(\epsilon_2+\epsilon_1+a_0\right)\\
b_2 & = & {{\left(\epsilon_2-\epsilon_1\right)\,s^2}\over{\epsilon_2\,\left(
 \epsilon_2+a_0\right)^2}}-{{\left(\epsilon_2-\epsilon_1\right)\,s^2
 }\over{\epsilon_1\,\left(\epsilon_1+a_0\right)^2}}+{{4\,\epsilon_1\,
 \epsilon_2^3+a_0\,\left(8\,\epsilon_1\,\epsilon_2^2+8\,\epsilon_1^2
 \,\epsilon_2\right)+12\,\epsilon_1^2\,\epsilon_2^2+4\,\epsilon_1^3\,
 \epsilon_2+b_0}\over{\epsilon_1\,\epsilon_2}}\\
b_1 & = & {{\left(\epsilon_2^2-\epsilon_1\,\epsilon_2\right)\,s^2}\over{
 \epsilon_1\,\left(\epsilon_1+a_0\right)^2}}-{{\left(\epsilon_1\,
 \epsilon_2-\epsilon_1^2\right)\,s^2}\over{\epsilon_2\,\left(
 \epsilon_2+a_0\right)^2}}-{{4\,\epsilon_1^2\,\epsilon_2^3+4\,
 \epsilon_1^3\,\epsilon_2^2+8\,a_0\,\epsilon_1^2\,\epsilon_2^2+b_0\,
 \epsilon_2+b_0\,\epsilon_1}\over{\epsilon_1\,\epsilon_2}}
\end{eqnarray*}
Once these constraints are implemented, we can write the Hamiltonians on the faces
of the image of the moment map, which are therefore parametrized by $a_0$ and $b_0$:
\begin{eqnarray*}
H_1&=&{{\left(\epsilon_2^2+\epsilon_1^2\right)\,s^2+a_0\,\left(8\,
 \epsilon_1^3\,\epsilon_2^2+8\,\epsilon_1^4\,\epsilon_2-2\,b_0\,
 \epsilon_1\right)+a_0^2\,\left(4\,\epsilon_1^2\,\epsilon_2^2+16\,
 \epsilon_1^3\,\epsilon_2-b_0\right)+4\,\epsilon_1^4\,\epsilon_2^2+8
 \,a_0^3\,\epsilon_1^2\,\epsilon_2-b_0\,\epsilon_1^2}\over{2\,
 \epsilon_1\,\epsilon_2^2-2\,\epsilon_1^2\,\epsilon_2}} 
 \\ &&
 -{{\epsilon_1
 \,s^2}\over{\epsilon_2\,\left(\epsilon_2+a_0\right)}}+{{\left(
 \epsilon_1\,\epsilon_2-\epsilon_1^2\right)\,s^2}\over{2\,\epsilon_2
 \,\left(\epsilon_2+a_0\right)^2}}+{{s^2}\over{\epsilon_1+a_0}}
 \end{eqnarray*}

\begin{eqnarray*}
H_2&=&-{{\left(\epsilon_2^2+\epsilon_1^2\right)\,s^2+a_0\,\left(8\,
 \epsilon_1\,\epsilon_2^4+8\,\epsilon_1^2\,\epsilon_2^3-2\,b_0\,
 \epsilon_2\right)+4\,\epsilon_1^2\,\epsilon_2^4+a_0^2\,\left(16\,
 \epsilon_1\,\epsilon_2^3+4\,\epsilon_1^2\,\epsilon_2^2-b_0\right)+8
 \,a_0^3\,\epsilon_1\,\epsilon_2^2-b_0\,\epsilon_2^2}\over{2\,
 \epsilon_1\,\epsilon_2^2-2\,\epsilon_1^2\,\epsilon_2}}
  \\ &&
 +{{s^2}\over{
 \epsilon_2+a_0}}
 -{{\left(\epsilon_2^2-\epsilon_1\,\epsilon_2\right)
 \,s^2}\over{2\,\epsilon_1\,\left(\epsilon_1+a_0\right)^2}}-{{
 \epsilon_2\,s^2}\over{\epsilon_1\,\left(\epsilon_1+a_0\right)}}
  \end{eqnarray*}

\begin{eqnarray*}
H_3&=&{{\left(\epsilon_2-\epsilon_1\right)\,s^2}\over{4\,\epsilon_2\,
 \left(\epsilon_2+a_0\right)^2}}-{{\left(\epsilon_2-\epsilon_1\right)
 \,s^2}\over{4\,\epsilon_1\,\left(\epsilon_1+a_0\right)^2}}-{{a_0\,
 \left(8\,\epsilon_1\,\epsilon_2^2+8\,\epsilon_1^2\,\epsilon_2\right)
 +4\,\epsilon_1^2\,\epsilon_2^2+12\,a_0^2\,\epsilon_1\,\epsilon_2-b_0
 }\over{4\,\epsilon_1\,\epsilon_2}}
  \end{eqnarray*}
 Remark that $b_0$ enters the formulae linearly so that the rank two faces are ruled surfaces.
To find the intersection between two faces, we go back to eq.(\ref{Q6R1}). As we have seen, the discriminant 
$b_1^2-4 b_0 b_2= -64\; \bar{b} b  $ is zero at a critical point and negative as soon as we leave this critical point. 
So the double real root at a critical point splits into a pair of complex conjugate roots as soon as we move along the rank $1$ line.
If the other discriminant  $a_1^2-4 a_0=(1/4)\Delta $ is positive, 
the polynomial $\lambda^{2}+a_{1}\lambda+a_{0}$ has real roots $\alpha_1$ and $\alpha_2$, 
and we can recast $Q_6$ as $Q_6(\lambda)=(\lambda-\alpha_1)^{2}((\lambda-\alpha_2)^{2}
(\tilde{b}_2\lambda^{2}+\tilde{b}_1\lambda+\tilde{b}_{0}))$. 
Expanding the second factor (of degree four), we obtain an expression of
the form (\ref{Q6R2}) with {\it real} coefficients, which satisfies all the constraints. This establishes that the line of rank one,
defined by eq.(\ref{Q6R1}) is included in a face of rank two, as long as the roots $\alpha_1$ and $\alpha_2$ are real, that is when
$\Delta>0$. When $\Delta$ crosses zero to become negative, the line of rank one starts to leave the face and goes 
inside the image of the moment map. The corresponding roots form a complex conjugated pair of double roots.  

We have also established that the points of rank zero are on the faces as well as the edges of rank one, provided $\Delta >0$. 
This last condition  excludes the lines of rank 1 which go in the interior of the image of the moment map when $\Delta < 0$. 
 
 \begin{figure}[h]
 \vskip -4cm
\begin{center}
\includegraphics[height= 25cm]{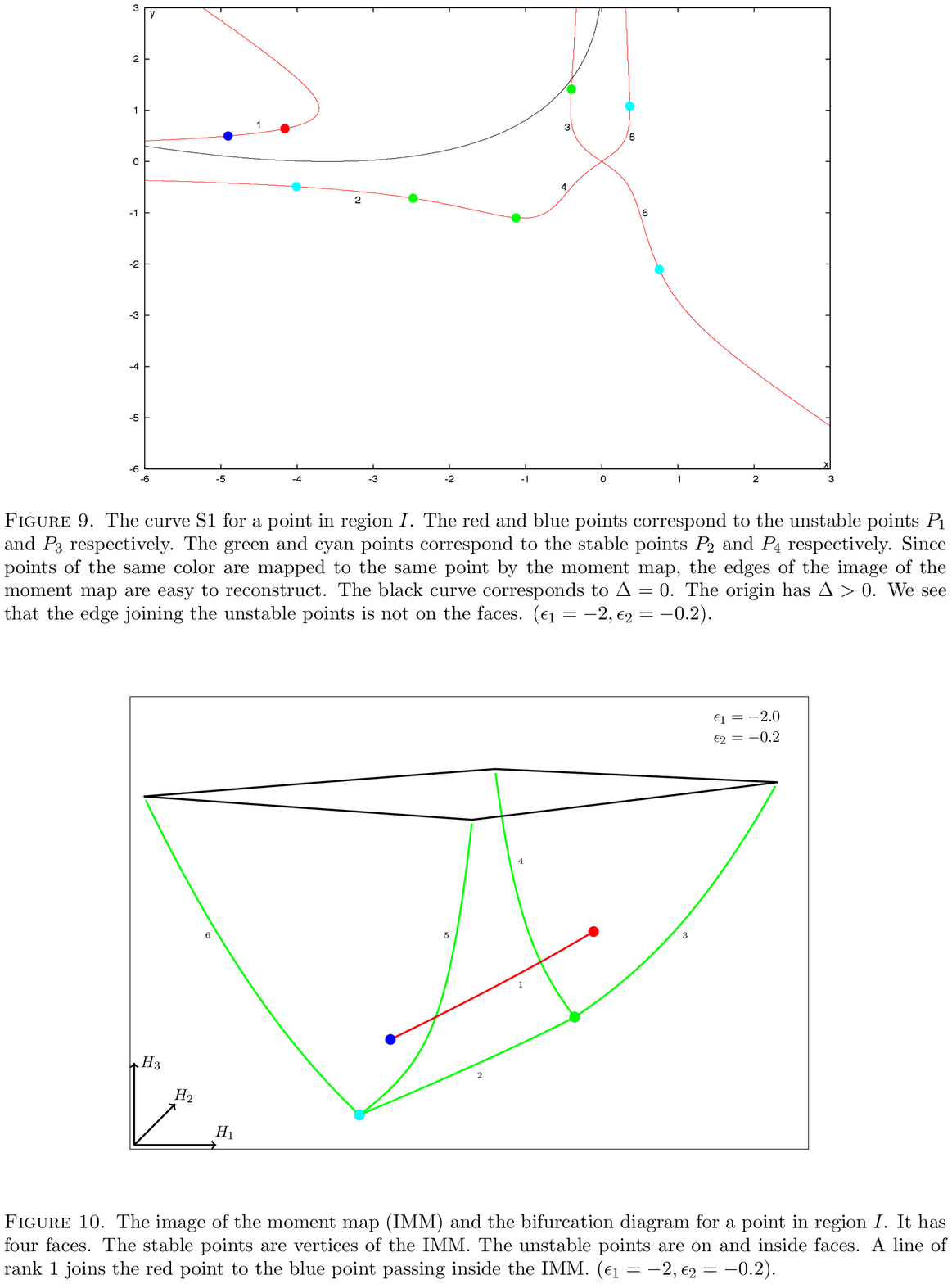}
\label{range1}
\end{center}
\nonumber
\end{figure}

% \begin{figure}[h]
%\begin{center}
%\includegraphics[width= 14.5cm]{S1_rangeI}
%\caption{The curve $S_1$ for a point in region $I$. The red and blue points correspond to the unstable points $P_1$ and $P_3$ respectively. The green and cyan points correspond to the stable points $P_2$ and $P_4$ respectively.
%Since points of the same color are mapped to the same point by the moment map, 
%the edges of the image of the moment map are easy to reconstruct. 
%The black curve corresponds to $\Delta=0$. The origin has $\Delta>0$. 
%We see that the edge joining the unstable points is not on the faces.  $(\epsilon_1=-2, \epsilon_2= -0.2)$.}
%\label{range1}
%\end{center}
%\nonumber
%\end{figure}
%
% \begin{figure}[h]
%\begin{center}
%\includegraphics[width= 14.5cm]{polytope2spins_I}
%\caption{The image of the moment map (IMM) and the bifurcation diagram for a point in region $I$.  It has four faces. 
%The stable points are vertices of the IMM. The unstable points are on and inside faces. 
%A line of rank $1$ joins the red point to the blue point passing inside the IMM.  $(\epsilon_1=-2, \epsilon_2= -0.2)$.}
%\label{polytope1}
%\end{center}
%\nonumber
%\end{figure}

 \begin{figure}[h]
 \vskip -4cm
\begin{center}
\includegraphics[height= 25cm]{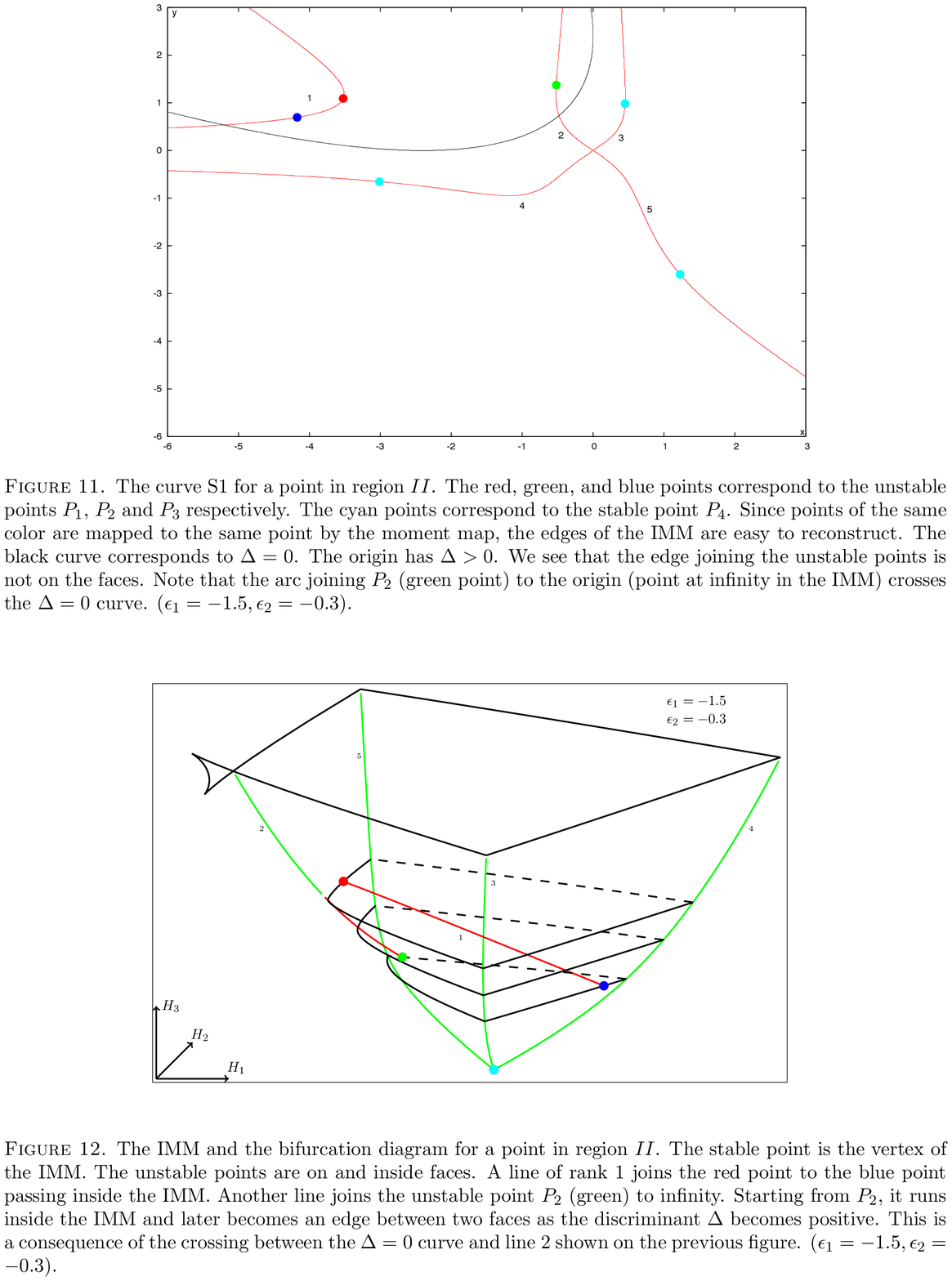}
\label{range2}
\end{center}
\nonumber
\end{figure}

% \begin{figure}[h]
%\begin{center}
%\includegraphics[width= 13.5cm]{S1_rangeII}
%\caption{The curve $S_1$ for a point in region $II$. The red, green, and blue points correspond to the unstable points 
%$P_1$, $P_2$ and $P_3$ respectively. The  cyan points correspond to the stable point  $P_4$.
%Since points of the same color are mapped to the same point by the moment map, the edges of the IMM are easy to reconstruct. 
%The black curve corresponds to $\Delta=0$. The origin has $\Delta>0$. 
%We see that the edge joining the unstable points is not on the faces.  
%Note that the arc joining $P_2$ (green point) to the origin (point at infinity in the IMM) crosses the  $\Delta=0$ curve.
%$(\epsilon_1=-1.5, \epsilon_2= -0.3)$.}
%\label{range2}
%\end{center}
%\nonumber
%\end{figure}
%
% \begin{figure}[h]
%\begin{center}
%\includegraphics[width= 13.5cm]{polytope2spins_II}
%\caption{The IMM and the bifurcation diagram  for a point in region $II$. The stable point is the vertex of the IMM. The unstable points are on and inside faces. 
%A line of rank $1$ joins the red point to the blue point passing inside the IMM. Another line joins the 
%unstable point $P_2$ (green) to infinity. Starting from $P_2$, it runs inside the IMM and later becomes an edge between
%two faces as the discriminant $\Delta$ becomes positive. This is a consequence of the crossing between the $\Delta=0$ curve
%and line 2 shown on the previous figure.  $(\epsilon_1=-1.5, \epsilon_2= -0.3)$.}
%\label{polytope2}
%\end{center}
%\nonumber
%\end{figure}

 \begin{figure}[h]
 \vskip -4cm
\begin{center}
\includegraphics[height= 25cm]{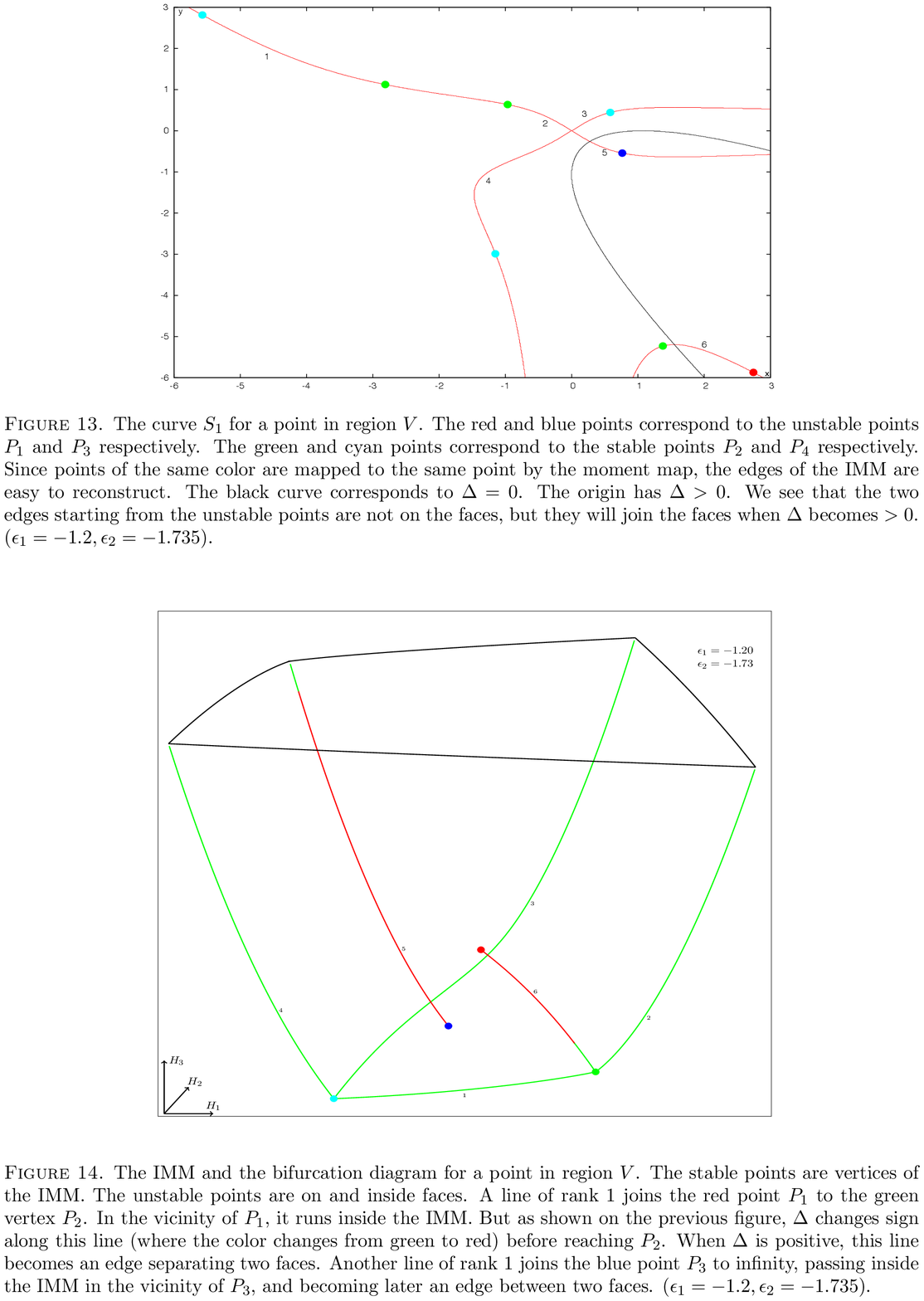}
\label{range5}
\end{center}
\nonumber
\end{figure}

\section{Conclusion}
In this article we have analyzed the bifurcation diagram of the Jaynes-Cummings model. The use of Lax pair techniques proved to be very useful. The classical analogue of algebraic Bethe Ansatz  allows a very easy construction of the normal forms near critical points.
We have shown in this way that this model possesses  singularities of the elliptic and focus-focus type only. This is an approach alternative to the one due to Krichever \cite{Krich83} and based on the  spectral curve. The spectral curve however was a very  powerful tool to  draw the full bifurcation diagram as advocated by Mich\`ele Audin \cite{Audin96}. In the one spin case we get results in agreement with general considerations (\cite{San10,Zung97}), while in the two spins case it exhibits quite a rich structure which calls for more detailed investigations.
 
 Along the open questions is the determination of normal forms along the lines of rank one, and the explicit construction of {\em real} solutions of the equations of motion along these lines. Another fascinating subject is the emergence of this classical geometry from the quantum system and Bethe equations~\cite{BaTal07}. 
We hope to return to these questions in future publications.


\begin{thebibliography}{XXXXX}
\bibitem{williamson36} John Williamson, {\it On the Algebraic Problem Concerning the Normal Forms of Linear Dynamical Systems.} American Journal of Mathematics, Vol. 58 No. 1 (1936), pp. 141-163.

\bibitem{Arnold97} V. I. Arnold, {\it Mathematical Methods of Classical Mechanics}, Springer, New-York, 1997, Appendix 6.

\bibitem{Eliasson90} L. H. Eliasson {\it Normal forms for Hamiltonian systems with Poisson commuting
integrals - elliptic case} Comment. Math. Helvetici 65 (1990) pp. 4-35.

\bibitem{Dicke} R. H. Dicke, {\it Coherence in spontaneous radiation processes}, Phys. Rev. {\bf 93}, 99 (1954).

\bibitem{JC} E. Jaynes, F. Cummings, Proc. IEEE vol. 51 (1963) p. 89.

\bibitem{Gau83} M. Gaudin, {\bf La Fonction d' Onde de Bethe.}  Masson, (1983).

\bibitem{YKA} E.  Yuzbashyan, V. Kuznetsov, B. Altshuler, {\it Integrable dynamics of coupled Fermi-Bose condensates.}
Phys. Rev. B {\bf 72} (2005), p. 144524.

\bibitem{Audin96} M. Audin, {\it Spinning tops}, Cambridge University Press 1996.

\bibitem{Atiyah82} M. F. Atiyah, {\it Convexity and commuting Hamiltonians.}  Bull. London Math. Soc. 14(1) (1982) pp. 1-15.

\bibitem{GuilStern82} V. Guillemin, S. Sternberg, {\it Convexity properties of the momentum mapping.} Invent. Math. 67(3) (1982) pp. 491-513.

\bibitem{San05} San V\~u Ngoc {\it Moment polytopes for symplectic manifolds with monodromy.} Adv. Math. {\bf 208} (2007), pp. 909-934

\bibitem{Duistermaat80} H. Duistermaat, {\it On Global Action-Angle variables}, Comm. Pure Appl. Math. 33 (1980) pp. 687-706.

\bibitem{San10} Alvaro Pelayo,  San V\~u Ngoc. {\it Hamiltonian dynamics and spectral theory for spin-oscillators.} 
ArXiv 1005.0439.

\bibitem{BaDoCa09} O. Babelon, B. Dou\c{c}ot,  L. Cantini, {\it A semiclassical study of the Jaynes-Cummings model.}   J. Stat. Mech. (2009) P07011.

\bibitem{Krich83} I. Krichever, {\it ``Hessians'' of integrals of the Korteweg-De Vries Equation and Perturbations of Finite-Zone Solutions.} Soviet Math. Dokl. Vol. 27 (1983), No. 3, pp. 757-761. 

\bibitem{Audin01} M. Audin, {\it Hamiltonian Monodromy via Picard-Lefschetz theory.} Comm. Math. Phys. 229 (2002) pp. 459-489.

\bibitem{Zou92}  M. Zou, {\it Monodromy in two degrees of freedom integrable systems}, J. Geom. Phys.{\bf 10}, (1992) p. 37.

\bibitem{Zung97} N. T. Zung, {\it A note on focus-focus singularities}, Lett. Math. Phys.  {\bf 60}, (2002), pp. 87-99.

\bibitem{Zung02} N. T. Zung, {\it Another note on focus-focus singularities}, Diff. Geom. Appl. {\bf 7}, (1997), p. 123.

\bibitem{Cushman01} R. Cushman and J. J. Duistermat, {\it Non-Hamiltonian monodromy}, J. Diff. Eqs. {\bf 172}, (2001) p. 42.

\bibitem{DuKrNo90} B.A. Dubrovin, I.M. Krichever, S.P. Novikov, {\it Integrable Systems I.} Encyclopedia of Mathematical Sciences, 
Dynamical systems IV. Springer (1990) p.173--281.

\bibitem{BaBeTa03} O. Babelon, D. Bernard, M. Talon, {\bf Introduction to Classical Integrable systems.}
Cambridge University Press (2003).

\bibitem{BaTa03} O. Babelon, M. Talon, {\it Riemann surfaces, separation of variables and classical and quantum integrability.} Phys. Lett. A. 312 (2003), pp. 71-77.

\bibitem{Sk79} E. Sklyanin, {\it Separation of variables in the Gaudin model.} J. Soviet Math., Vol. 47,
(1979) pp. 2473-2488.

\bibitem{Griffiths78} P. Griffiths and J. Harris, {\it Principles of algebraic geometry}, Wiley, New-York (1978), chapter 2.

\bibitem{BaTal07} O. Babelon, D. Talalaev, {\it On the Bethe Ansatz for the Jaynes-Cummings-Gaudin model.} 
J. Stat. Mech. (2007) P06013.

\end{thebibliography}
 \end{document}